\documentclass[prd,preprintnumbers,floatfix,
nofootinbib,superscriptaddress]{revtex4}
\usepackage{float}

\usepackage{nicefrac}
\usepackage{mathtools}
\usepackage{amsfonts} % AMS
\usepackage{amssymb} % AMS
\usepackage{amsmath} % AMS
\usepackage{graphicx} % Include figure files
\usepackage{subfigure} % Include figure files
\usepackage{array} % array
\usepackage{dcolumn} % Align table columns on decimal point
\usepackage{bm} % bold math
\usepackage{latexsym} % latex symbols
\usepackage{esint}
\usepackage{longtable} % long tables
\usepackage{hyperref} % hypertext links 
\usepackage{verbatim}
\usepackage{epsfig}
\usepackage{slashed}
\usepackage{color}
\usepackage{cancel}
\usepackage{soul}
\usepackage[multidot]{grffile}

\definecolor{cardinal}{rgb}{0.77, 0.12, 0.23}

\newcommand{\nn}[0]{\nonumber}
\newcommand{\cK}[0]{\mathcal K}
\newcommand{\cM}[0]{\mathcal M}

\newcommand{\cD}[0]{\mathcal D}

\newcommand{\diff}[0]{\mathrm{d}}

\newcommand{\p}{\mathbf{p}} 
\renewcommand{\k}{\mathbf{k}}

\newcommand{\df}[0]{\mathrm{df}}

\begin{document}
\title{
Solving relativistic three-body integral equations in the presence of bound states}

\newcommand{\jlab}{Thomas Jefferson National Accelerator Facility, 
12000 Jefferson Avenue, Newport News, Virginia 23606, USA}
\newcommand{\odu}{Department of Physics, 
Old Dominion University, 
Norfolk, Virginia 23529, USA}

%%%%%%%%%%%%%%%%%%%%%
\author{Andrew W. Jackura}
\email[e-mail: ]{ajackura@odu.edu}
\affiliation{\jlab}
\affiliation{\odu}
%%%%%%%%%%%%%%%%%%%%%

%%%%%%%%%%%%%%%%%%%%%
\author{Ra\'ul A.~Brice\~no}
\email[e-mail: ]{rbriceno@jlab.org}
\affiliation{\jlab}
\affiliation{\odu}
%%%%%%%%%%%%%%%%%%%%%

%%%%%%%%%%%%%%%%%%%%%
\author{Sebastian M. Dawid}
\email[e-mail: ]{sdawid@iu.edu}
\affiliation{Physics Department, Indiana University, Bloomington, Indiana 47405, USA}
\affiliation{Center for Exploration of Energy and Matter, Indiana University, Bloomington, Indiana 47403, USA}
%%%%%%%%%%%%%%%%%%%%%

%%%%%%%%%%%%%%%%%%%%%
\author{Md Habib E Islam}
\email[e-mail: ]{m2islam@odu.edu}
\affiliation{\odu}
%%%%%%%%%%%%%%%%%%%%%

%%%%%%%%%%%%%%%%%%%%%
\author{Connor McCarty}
\email[e-mail: ]{connormc757@gmail.com}
\affiliation{Matthew Fontaine Maury High School, Norfolk, Virginia 23517, USA}
%%%%%%%%%%%%%%%%%%%%%

%%%%%%%%%%
\date{\today}
%%%%%%%%%%

%%%%%%%%%%%%%%%%%%%%%%%%%%%%%%%%%%%%
%	Preprint Number
%%%%%%%%%%%%%%%%%%%%%%%%%%%%%%%%%%%%
\preprint{JLAB-THY-20-3272}

%%%%%%%%%%
\begin{abstract}
We present a systematically improvable method for numerically solving relativistic three-body integral equations for the partial-wave projected amplitudes. The method consists of a discretization procedure in momentum space, which approximates the continuum problem with a matrix equation. It is solved for different matrix sizes, and in the end, an extrapolation is employed to restore the continuum limit. Our technique is tested by solving a three-body problem of scalar particles with an $S$ wave two-body bound state. We discuss two methods of incorporating the pole contribution in the integral equations, both of them leading to agreement with previous results obtained using finite-volume spectra of the same theory. We provide an analytic and numerical estimate of the systematic errors. Although we focus on kinematics below the three-particle threshold, we provide numerical evidence that the methods presented allow for determination of amplitude above this threshold as well.
\end{abstract}
%%%%%%%%%%

%\keywords{weak decays, lattice QCD}

\nopagebreak

\maketitle

%%%%%%%%%%%%%%%%%%%%%%%%%%%%%%%%%%%%
%	Section :: Introduction
%%%%%%%%%%%%%%%%%%%%%%%%%%%%%%%%%%%%
\section{Introduction}
\label{sec:intro}

Several outstanding problems in modern-day hadronic, particle, and nuclear physics require a relativistic description of the dynamics of multi-hadron systems. Many resonances, which challenge our understanding of the strong interaction, are observed experimentally in reactions involving final states composed of three particles or more. One example is the recently observed tetraquark candidate $X(2900)$ found in the $B^+\to D^+D^-K^+$ decay \cite{Aaij:2020hon, Aaij:2020ypa}. Due to the complexity of these reactions, it is rarely evident if these are indeed genuine states of quantum chromodynamics (QCD), or merely kinematic enhancements \cite{Burns:2020epm, Burns:2020xne}. Similarly, three-body decays play a significant role in modern-day tests of the fundamental symmetries of the Standard Model and searches of its extensions. A prominent example is the measurement of the enhanced CP violations in $B^\pm$ decays to three light mesons~\cite{Aaij:2014iva}, where the large CP asymmetries can result from the presence of a rich resonant structure in the three-body final state.  Lastly, it is well known that the three-nucleon forces are indispensable in the effective description of light nuclei and their properties [see Refs.~\cite{Mermod:2004, Mermod:2005, Piarulli:2017dwd} for some key examples]. However, the exact form of their contribution, within the context of QCD, is still undetermined, see Ref.~\cite{Ekstrom:2020}.

To resolve these, and many other problems, a coordinated effort has been initiated to obtain two- and three-hadron dynamics from QCD using lattice QCD.\footnote{For recent reviews on this topic we point the reader to Refs.~\cite{Briceno:2017max, Hansen:2019nir}.} Although the scattering amplitudes are not accessible directly in the finite volume computations, one can obtain them from the finite-volume spectra computed with lattice QCD via appropriate non-perturbative mappings called {\it quantization conditions}~\cite{Luscher:1986pf,Luscher:1990ux}. This technique has proven successful in the two-hadron sector~\cite{Dudek:2010ew,Beane:2011sc,Pelissier:2012pi,Dudek:2012xn,Liu:2012zya,Beane:2013br,Orginos:2015aya,Berkowitz:2015eaa,Lang:2015hza,Bulava:2016mks,Hu:2016shf,Alexandrou:2017mpi,Bali:2017pdv,Bali:2017pdv,Wagman:2017tmp,Andersen:2017una,Brett:2018jqw,Werner:2019hxc,Mai:2019pqr,Wilson:2019wfr,Cheung:2020mql, Rendon:2020rtw}, including systems where multiple open channels are kinematically accessible~\cite{Wilson:2014cna,Dudek:2014qha,Wilson:2015dqa,Dudek:2016cru,Briceno:2016mjc,Moir:2016srx,Briceno:2017qmb,Woss:2018irj,Woss:2019hse, Woss:2020ayi}. Extensions of this methodology to the three-particle sector have been formally developed in recent years, focusing on three identical scalar particles.
Two approaches have been followed to address the determination of relativistic three-particle scattering amplitude from lattice QCD. The first is the relativistic field theory (RFT) approach, which derives the quantization condition by summing on-shell projected generalized Feynman diagrams to all-orders~\cite{Hansen:2014eka, Hansen:2015zga, Hansen:2016, Briceno:2017, Briceno:2018, Briceno:2018a, Briceno:2019muc, Blanton:2019a, Hansen:2020, Blanton:2020gha}.
An alternative method based on $S$ matrix unitarity, called the finite volume unitarity (FVU) approach, constructs on-shell scattering equations from the unitarity relation for the amplitude~\cite{Mai:2017vot,Jackura:2018xnx,Mikhasenko:2019vhk, Dawid:2020uhn}, then postulates a finite volume analog which can be used to derive a quantization condition~\cite{Mai:2017bge,Doring:2018xxx,Mai:2018djl}. These methods have been shown to result in equivalent infinite volume scattering equations~\cite{Jackura:2019bmu} and quantization conditions~\cite{Blanton:2020jnm}. Both approaches introduce an unknown function which describes the short-distance three-body interactions, which is to be determined from lattice QCD. To be concrete, we follow the RFT approach, where this function is denoted $\cK_{\text{df},3}$. This three-body $K$ matrix feeds into a set of integral equations, which when solved yield an on-shell representation for the three-particle scattering amplitude. Recently, the first applications of these formalisms have been used to determine the interactions of $3\pi^+$~\cite{Horz:2019rrn,Blanton:2019vdk,Mai:2019fba,Culver:2019vvu,Fischer:2020jzp,Hansen:2020otl}, as well as $3K^+$~\cite{Alexandru:2020xqf}. The most recent generalization of the RFT formalism incorporates all possible values of two- and three-pion isospin~\cite{Hansen:2020zhy}, therefore allowing the more difficult cases to be studied, e.g. $a_1 \to \pi \rho \to 3\pi$, scattering in the $S$- and $D$-wave channels, see Refs.~\cite{Mikhasenko:2018bzm,Sadasivan:2020syi} for recent investigations of this channel. For such systems, one needs to define carefully the procedure of solving the three-body equations numerically to arrive at reliable results.

Recently, Ref.~\cite{Hansen:2020otl} presented a framework for evaluating these integral equations for weakly-interactive systems, i.e. for small $\cK_{\text{df},3}$ and weak coupling between particles in the two-body sub-channels. It was used to analyze the $3\pi^+$ finite-volume spectrum obtained via Lattice QCD, resulting in the first three-particle energy-dependent scattering amplitude from QCD. In this work, we investigate numerical solutions to the integral equations describing relativistic three-body systems presented in Refs.~\cite{Hansen:2015zga, Mai:2017a, Jackura:2018xnx}. We study a more challenging case than the one presented in Ref.~\cite{Hansen:2020otl}, namely, we consider the scattering of three scalar particles in the presence of two-body bound states.  This can be characterized as a toy model of three-nucleon systems where the deuteron, as a shallow bound state in the two-nucleon sector, can be formed.

Solutions are obtained by introducing a discretized mesh in momentum space to numerically approximate the integral equations by a system of $N$ linear equations. The presence of the two-body bound state results in a pole singularity in the integration range, which can lead to numerical instabilities. We consider two regularization methods, both relying on introducing a finite $\epsilon$ to avoid the pole as the momentum sweeps over the kinematic region. One method introduces the regulator as the usual $+i\epsilon$ prescription to move the pole off the real axis, while the other removes the imaginary part of this pole with an $\epsilon$-regulated delta function. Solutions of the matrix equation are extrapolated to the continuum by a careful investigation of the discrete mesh parameters $N$ and $\epsilon$ in their appropriate limits. 

To test the quality of our results, we impose several tests to assess the systematic effects of our solution strategy. Our primary assessment comes in the form of testing our solution against the unitarity constraints of two-body scattering below the three-body threshold. Additionally, we investigate the convergence of our solutions against variations in the meshing parameters and extrapolation procedure, finding that we can recover solutions to sub-percent level deviation from $S$ matrix unitarity. Finally, below the three-particle threshold, we find an agreement with an independent numerical study of the same toy model using the corresponding finite volume formalism presented in Ref.~\cite{Romero-Lopez:2019qrt}, which provides further confidence in our findings. Since the method introduced here works in the most demanding scenario, we claim that the presented framework applies to a general three-body system and provides a systematic procedure for obtaining solutions to the integral equations.

This work is organized in the following way. First, in Sec.~\ref{sec:IE}, we review the theoretical three-body framework of interest and define the integral equations to be solved. Next, in Sec.~\ref{sec:numerics}, we present the solutions of these equations, compare them with other approaches, and discuss different physical scenarios under consideration. In Sec.~\ref{sec:details}, we explain the details of two numerical methods used to obtain the results. In Sec.~\ref{sec:syst}, systematic errors of those methods are described. We discuss the extrapolation of the discrete solutions to the continuum limit and their independence on the meshing procedure. In Sec.~\ref{sec:above_thr} we present and briefly comment on the solutions in the region above the three-body threshold. Finally, in Sec.~\ref{sec:conclusions}, we provide a summary of our work.

%%%%%%%%%%%%%%%%%%%%%%%%%%%%%%%%%%%%
%	Section :: Integral equations
%%%%%%%%%%%%%%%%%%%%%%%%%%%%%%%%%%%%
\section{Integral equations}
\label{sec:IE}

The integral equations describing relativistic three-body amplitudes, which are the main focus of this work, were presented in Refs.~\cite{Hansen:2015zga, Mai:2017a, Jackura:2018xnx}. Here, we follow the prescription of Ref.~\cite{Hansen:2015zga}. There, in the original formulation of the equations, it was necessary to assume that there are no narrow resonances or bound states in the two-body sub-channels of the general three-body states. Reference~\cite{Romero-Lopez:2019qrt} showed that this assumption can be lifted by a simple modification of the two-body phase space. As will be seen, for the scenario considered here, the equations of Ref.~\cite{Hansen:2015zga} are explicitly unchanged. This is to be expected since Ref.~\cite{Jackura:2018xnx} derived an equivalent set of integral equations using unitarity and did not make any assumptions about the form of the two-body dynamics.

\begin{figure}[t!]
        \centering
        \includegraphics[ width=0.75\textwidth]{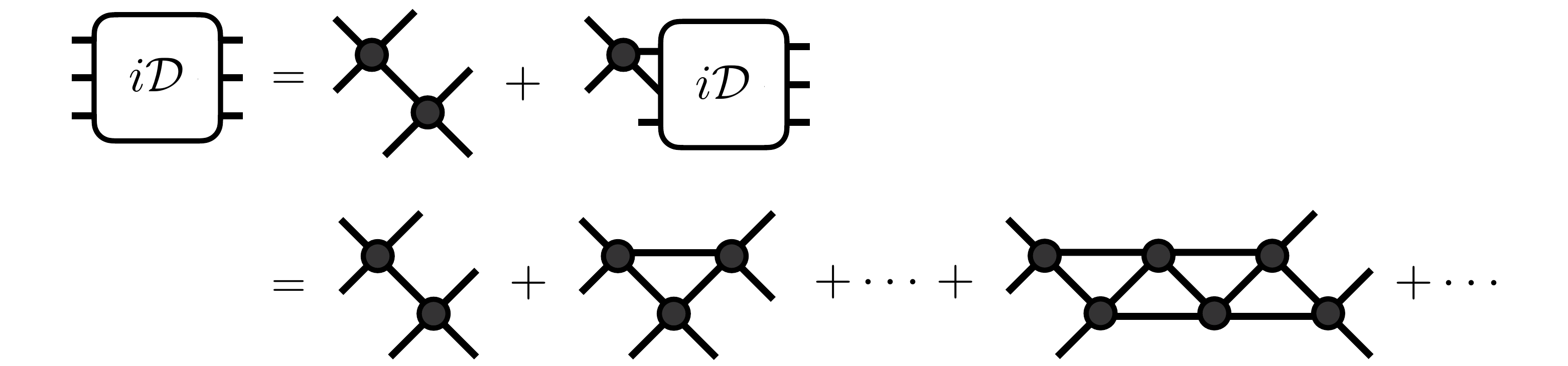}
        \caption{Diagrammatic representation of the $\cD^{(u,u)}$ amplitude defined in Eq.~\eqref{eq:Ddef}. Black circles represent the on-shell $2\to 2$ amplitude $\cM_2$.}
        \label{fig:ladder}
    \end{figure}

After these preliminary comment, we proceed to review the integral equations of interest. The unsymmetrized three-body scattering amplitude can be written as:
    %%%%%
    \begin{align}
    \label{eq:Def.M3uu}
    \cM^{(u,u)}_3( \p,\k) = \cD^{(u,u)}(\p, \k) + \cM_{\df,3}^{(u,u)}( \p,\k ),
    \end{align}
    %%%%%
where $\cD^{(u,u)}$, called the \emph{ladder} amplitude depicted in Fig.~\ref{fig:ladder}, contains the sum over all possible pair-wise interactions via a sequence of one-particle exchanges and $\cM^{(u,u)}_{\df,3}$ is driven by the three-body $K$ matrix $\cK_{\df,3}$ representing short-distance three-particle interactions. Here $\k$ and $\p$ are the momenta of one of the hadrons in the initial and final state, respectively. We refer to this hadron as the \emph{spectator}. The other two hadrons, called a \emph{pair}, associated with the given spectator are projected to definite angular momentum. Since only total angular momentum is conserved, the unsymmetrized amplitude is a non-diagonal matrix in the pair's angular momentum space. In addition to the external momentum and angular dependencies, the amplitude depends on the total center-of-momentum (CM) energy of the three-particle system, denoted by $E$, which is suppressed in the argument list of Eq.~\eqref{eq:Def.M3uu}. 

From here on we make two simplifications. First, we assume that the two-body subsystem contains contributions from the $\ell = 0$ partial wave only, leaving just one matrix element of Eq.~\eqref{eq:Def.M3uu} for our consideration. The method of the numerical solution presented here applies to any partial wave amplitude, but we constrain ourselves to the $S$ wave case just for simplicity. Although here we are primarily interested in the unsymmetrized amplitude, the fully symmetrized amplitude is obtained by summing over the nine possible spectator momenta, which for pairs in $S$ wave reduces to
    %%%%%
    \begin{equation}
    \label{eq:symm_proc}
    \mathcal M_3(E; \p, \mathbf{a}' ;  \k, \mathbf{a} ) =  \sum_{\p \in \mathcal P_p }      \sum_{\k     \in \mathcal P_k } \cM^{(u,u)}_3(\p, \k) \,,
    \end{equation}
    %%%%%
where $\mathcal P_p = \{   \p, \mathbf{a}' ,  - \p -  \mathbf{a}'  \}$ and $\mathcal P_k = \{   \k, \mathbf{a} ,  - \k -  \mathbf{a}  \}$ and $\mathbf{a}$ and $\mathbf{a}'$ are the momenta of one of the particle in the initial and final pair state, respectively. The second simplification we describe below.

The first term appearing in Eq.~\eqref{eq:Def.M3uu} is $\cD$, which represents the sum over all possible pair-wise interactions mediated by one-particle exchanges, is defined by the integral equation (see Fig.~\ref{fig:ladder}),
    %%%%%
    \begin{equation}
    \label{eq:Ddef}
    \cD^{(u,u)}(\p, \k) = -\cM_2( p) G(\p, \k) \cM_2( k) - \cM_2( p) \int \! \frac{\diff^3 \k'}{(2 \pi)^3 2 \omega_{k'}}  \, G(\p, \k')   \cD^{(u,u)}(\k', \k)  \,.
    \end{equation}
    %%%%%
Here $\cM_2$ is the $S$ wave $2 \to 2$ scattering amplitude describing the initial and final interactions among particles in the pair. Their energy is fixed by the momentum of the spectator, $s_{2k}\equiv E_{2,k}^{\star 2} \equiv (E - \omega_k)^2 -  k^2 $, where $\omega_k \equiv \sqrt{m^2+k^2}$, $k \equiv \lvert \k \rvert$, and the $E_{2,k}^{\star}$ is the energy of the pair evaluated in their CM frame. The ladder amplitude $\cD$ is driven by the exchange propagator $G$, which describes the long-range interactions between the intermediate pair and spectator, and is defined as
    %%%%%
    \begin{equation}
    \label{eq:G}
    G(\p, \k) \equiv \frac{H(p, k)}{b_{pk}^2 - m^2 + i \epsilon} \,,
    \end{equation}
    %%%%%
where $b^2_{pk} \equiv (E - \omega_p - \omega_k)^2 - (\p + \k)^2 $, and $H( p, k)$ is a cut-off function to render the integral in Eq.~\eqref{eq:Ddef} finite. We use the cut-off defined in Ref.~\cite{Hansen:2015zga}, 
    %%%%%%
    \begin{align}
    \label{eq:cutoff}
    H(p, k) & \equiv J(E_{2,p}^{\star 2}/ 4 m^2 )  J(E_{2,k}^{\star 2}/ 4 m^2 ) \,, \\[5pt]
    %%%
    J(x) & \equiv
    \begin{cases}
    0 \,, & x \le 0 \\
    %%%
    \exp \left( - \frac{1}{x} \exp \left [-\frac{1}{1-x} \right] \right ) \,, 
    & 0<x \le 1 \\
    %%%
    1 \,, & 1<x \,,
    \end{cases}
    \end{align}
    %%%%%
such that the function evaluates to unity in the physical region and smoothly interpolates to zero at $E_{2,k}^{\star} = 0$.
    %%%%%%%%%%%%%
    % FIGURE
    %%%%%%%%%%%%%
    %%%%%%%%%%%%%

It is advantageous to work with the amputated amplitude $d$, which reduces the singularities associated with the initial and final scattering of the two-particle system associated with the pair. Following Ref.~\cite{Jackura:2018xnx}, we define it as: 
    %%%%%
    \begin{align}
    \label{eq:d_def}
    \cD^{(u,u)}(\p, \k) \equiv   \cM_2( p ) \, d^{(u,u)}(\p, \k) \, \cM_2( k) .
    \end{align}
    %%%%%%
It is natural to expect that it is less susceptible to instabilities when evaluated numerically close to poles of the two-body amplitudes $\cM_2$. It is important to emphasize that this is not an approximation, but a definition of $d$, which is more advantageous for systems where the two-body subsystems have either bound states or resonances. Using above equation and Eq.~\eqref{eq:Ddef}, we see that $d$ satisfies
    %%%%%
    \begin{equation}
    \label{eq:d_intdef}
    d^{(u,u)}(\p, \k) = - G(\p, \k)  -   \int \frac{\diff^3 \k'}{(2 \pi)^3 2 \omega_{k'}} \, G(\p, \k') \cM_2( k' \,) \, d^{(u,u)}(\k', \k)  \,.
    \end{equation}
    %%%%%

Equation~\eqref{eq:d_def} makes evident the claim that $d$ is less sensitive to singularities associated with the initial/final two-particle states. In particular, if these couple to bound states, $\mathcal{D}$ will have poles on the real axis while $d$ will not. Nevertheless, $d$ does depend on $\mathcal{M}_2$, and as a result $d$ can still exhibit singular behavior at the two-particle thresholds. In Sec.~\ref{sec:above_thr} we provide numerical evidence of the manifestation of these singularities. Furthermore, we give illustrative comparisons between $d$ and $\mathcal{D}$ for kinematics near the two-particle thresholds and a two-body bound state. 

The ladder amplitude does not contain any information about short-range three-body physics. This is described by the amplitude $\cM_{\df,3}$, which is the second term of Eq.~\eqref{eq:Def.M3uu}. The short-distance interactions are encoded into a three body $K$-matrix, denoted $\cK_{\df,3}$, which is the driving term for the integral equation for $\cM_{\df,3}$. This equation depends on $\cK_{\df,3}$ as well as the ladder amplitude $\cD$ (see for example the discussion in Sec. V of Ref.~\cite{Hansen:2015zga}). Therefore, within the framework of Ref.~\cite{Hansen:2015zga} one must determine $\cD$ first, and then for a given $\cK_{\df,3}$ the second integral equation for $\cM_{\df,3}$ can be solved. 

This bring us to the second simplification we make in this study. From this point forward, we assume that the three-body $K$ matrix is zero, thus the scattering amplitude is dominated by exchanges between two-particle subprocesses. Given this assumption, the explicit dependence of $\cM_{\df,3}$ on $\cK_{\df,3}$ is not needed here, only the fact that as $\cK_{\df,3} \to 0$, also $\cM_{\df,3} \to 0$, and therefore the three body amplitude is reduced to its ladder part:
    %%%%%
    \begin{align}
    \lim_{\cK_{\df,3} \to 0} \cM_3^{(u,u)}(\p,\k) = \cD^{(u,u)}(\p,\k) \, .
    \end{align}
    %%%%%
Our focus here is to develop an efficient framework for evaluating the integral equation for the ladder amplitude in the most singular scenario, namely when the two-body amplitude $\cM_2$ has a bound state pole present in the integration range of Eq.~\eqref{eq:Ddef}. As we have already emphasized, it is more efficient to determine the amputated amplitude of Eq.~\eqref{eq:d_def}, as in this case the external pole contributions are absent from the calculation, and thus in the remainder of this work we focus on $d$. Having determined $d$, including a non-zero $\cK_{\df,3}$ contribution is straightforward. This would, of course, require first determining $\cK_{\df,3}$ either from the lattice QCD spectrum~\cite{Hansen:2014eka} or experimental data.

To summarize, we consider two key approximations. The two-body subsystem is saturated by $\ell=0$ and the $\cK_{\df,3}=0$. This is the same limit considered by Ref.~\cite{Romero-Lopez:2019qrt}. As a result, in Sec.~\ref{sec:numerics}, we are able to provide direct comparison of the numerical solutions of the integral equations obtained here with those obtained in the aforementioned reference. The advantage of the techniques presented here are multifold. First, the numerical solutions can be reached with timeframes that are orders of magnitude shorter. This allows to map these functions continuously. Second, the solutions are systematically improvable. Third, the framework presented here holds for any kinematics above the three-particle thresholds, which is not the case for Ref.~\cite{Romero-Lopez:2019qrt}.

%%%%%%%%%%%%%%%%%%%%%%%%%%%%%%%%%%%%
%	Section :: The J = 0 scattering amplitude
%%%%%%%%%%%%%%%%%%%%%%%%%%%%%%%%%%%%
\subsection{The $J=0$ scattering amplitude}

The integral equation for $\cD$ (or for equivalently $d$) is taxing due to two main issues. First, the integrand is singular. This can be seen, for example, in Eq.~\eqref{eq:d_intdef} where the two-body amplitude $\cM_2$ appears under the integral, which in general can have branch cuts and poles. Secondly, the integral is three-dimensional, which makes the problem of the numerical solution of the equation much more complex.

We can avoid the second complication by employing the partial wave projection in total angular momentum $J$, which leads to an infinite number of one-dimensional integral equations that are simpler to evaluate. In practice for a system with definite quantum numbers, only a finite number of these equations must be solved. Since here we consider the limit where the two-particle subsystem has a single non-zero partial wave, namely $\ell =0$, the only source of angular dependence arises from the relative momentum between the two-particle subsystem and the spectator. In the CM frame, it is given by an angle of the spectator momentum. In other words, 
    %%%%%
    \begin{align}
    \label{eq:Def.D_PW}
    \cD^{(u,u)}(\p, \k) &=
    4\pi \sum_{J, m_J}\sum_{J', m_{J'}}  \, Y_{J m_J } (\hat{\p}) \,  \cD^{(u,u)}_{J m ; J'm_{J'}}( p,  k ) \,  Y_{J' m_{J'} }^{*} (\hat{\k}),
    \end{align}
    %%%%%
where $m_J$ is the projection onto some external $z$ axis. From angular momentum conservation, the resultant amplitude must be diagonal in $J$ and from azimuthal symmetry it should be independent of $m_J$,
    %%%%%
    \begin{align}
    \cD_{J m ; J'm_{J'}}^{(u,u)}(p,k) &=
    \delta_{JJ'} \delta_{m_{J}m_{J'}} \, \cD^{(u,u)}_{J }(p, k) \, .
    \end{align}
    %%%%%
For simplicity, we only consider the $J = 0$ component of the total amplitude, which is denoted by a subscript $S$ (for $S$-wave). With this, we can define the $J=0$ component of $\cD$ by integrating out the overall angle dependence of Eq.~\eqref{eq:Ddef} 
    %%%%%
    \begin{align}
    \label{eq:cD_Sproj}
     \cD^{(u,u)}_{S}(p, k) &\equiv
    \int \frac{\diff\Omega_{k}}{4\pi} \, \frac{\diff\Omega_{p}}{4\pi} \, \cD^{(u,u)}(\p, \k) \nn\\
    &=
    -\cM_2( p ) G_{S}(p, k) \cM_2( k ) - \cM_2( p) \int_0^\infty \! \frac{\diff k' \, k'^2 }{(2 \pi)^2 \,  \omega_{k'}} \, G_{S}(p,k') \cD^{(u,u)}_{S}( k',   k),
    \end{align}
    %%%%%
where we have introduced
    %%%%%
    \begin{align}
    \label{eq:Gs_proj}
    G_{S}(p, k)
    &\equiv \int \frac{\diff\Omega_{p}}{4\pi} \frac{\diff\Omega_{k}}{4\pi} \, 
    G(\p, \k)
    \nn\\[5pt]
    & = - \frac{H(p,k)}{4pk} \, \log\left( \frac{\alpha(p,k) - 2pk + i\epsilon}{\alpha(p,k) + 2pk + i\epsilon} \right)
    \end{align}
    %%%%%
where $\alpha(p,k) = (E-\omega_k - \omega_p)^2 - p^2 - k^2 - m^2$. We used the fact that $H$ is independent of the angle, see Eq.~\eqref{eq:cutoff}. By partial-wave projecting the exchange propagator, we have effectively softened its singularity from a pole to a logarithm. As one would expect, this further simplifies the numerical evaluation of the integral equations. We remark that having a non-zero value of $\epsilon$, was necessary to define the integral in Eq.~\eqref{eq:Gs_proj}. As we will see below, having a non-zero value of $\epsilon$ will play an important role in obtaining numerical solutions of the integral equations. Ultimately the desired solutions can be obtained by taking the $\epsilon\to 0$ limit of the subsequent solutions.

Since the initial and final state two-particle scattering amplitudes are unaffected by the angular projection performed above, the function $d$ defined by Eq.~\eqref{eq:d_def} has an expansion similar to Eq.~\eqref{eq:Def.D_PW}, and the integral equation for the partial wave projected $d$, Eq.~\eqref{eq:d_intdef}, reads
    %%%%%
    \begin{align}
    \label{eq:d_Sproj}
    d^{(u,u)}_{S}(p, k) 
    &=
    - G_{S}(p, k)   
    - \int_0^\infty \! \frac{\diff k' \, k'^2}{ (2\pi)^2 \, \omega_{k'}} \, G_{S}(p, k') \, \cM_2( k') \, d^{(u,u)}_{S}(k', k) \, .
    \end{align}
    %%%%%
In the remainder of this work, we consider this form of the ladder equation.

Equation Eq.~\eqref{eq:cD_Sproj} was solved in Ref.~\cite{Hansen:2020otl} to obtain the first three-body amplitudes from lattice QCD. Formally, successive iterations of Eq.~\eqref{eq:Ddef} yield a solution in terms of an increasing number of exchanges between the two-particle subsystems. For weakly coupled two-body systems, i.e. for small $ma$, the series rapidly converges and the first few orders dominate the solution. In the case considered here, i.e. for strongly interacting systems in which the two-particle subsystem forms a bound state, the perturbation series fails to converge and we are forced to resort to a non-perturbative numerical approach.

%%%%%%%%%%%%%%%%%%%%%%%%%%%%%%%%%%%%
%	Section :: Integral equations in the presence of a two body bound state
%%%%%%%%%%%%%%%%%%%%%%%%%%%%%%%%%%%%
\subsection{Integral equations in the presence of a two-body bound state}
\label{sec:IE_BS}

\begin{figure}[t!]
        \centering
        \includegraphics[ width=0.75\textwidth]{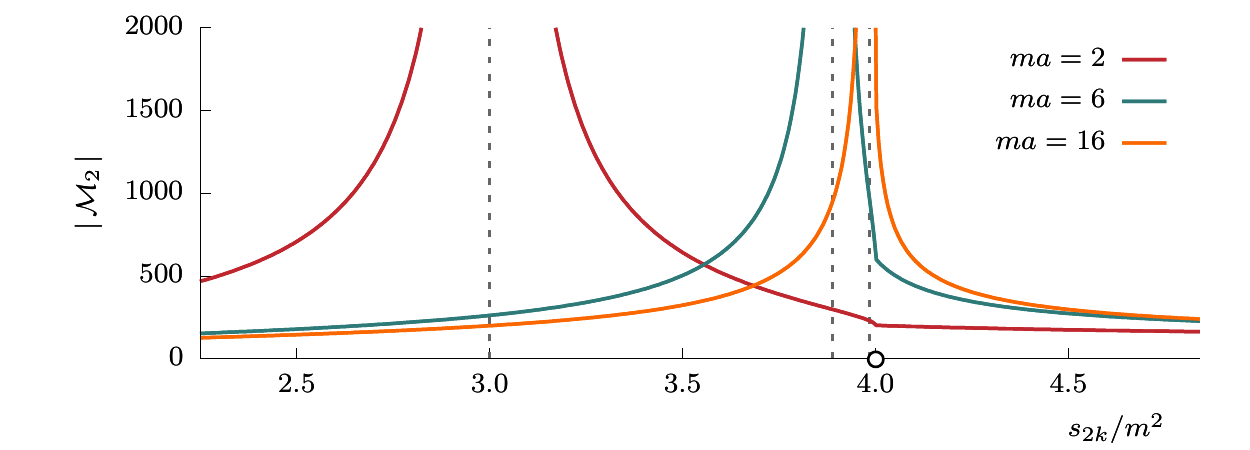}
        \caption{Plot of $\lvert\, \cM_2 \, \rvert$ as a function of $E_{2k}^{\star} / m$ for cases $ma = 2$, 6, and 16. The vertical dashed lines indicate locations of bound state poles $\sqrt{s_{b}} / m$, and the open circle on the $x$ axis corresponds to the threshold.}
        \label{fig:M2}
    \end{figure}

As announced in Sec.~\ref{sec:intro}, here we are interested in the implications of these integral equations for systems where the two-body subsystem can become bound. In this work, we consider the effective range expansion for $\cM_2$, and assume it is dominated by the leading order (LO) order term 
    %%%%%
    \begin{align}
    \label{eq:ERE}
    \cM_2( k) &= \cM_2(s_{2k}) \, ,\nn\\[5pt]
    %%%
    &= \frac{16 \pi E_{2,k}^{\star}  }{-1/a - i q_{2k}^{\star} } \,,
    \end{align}
    %%%%%
where $q_{2k}^{\star}=\sqrt{E_{2,k}^{\star 2} /4-m^2}$ is the relative momentum between the two particles in their CM frame and $a$ is the scattering length. Figure~\ref{fig:M2} shows plots of $\lvert\cM_2\rvert$ as a function of $E_{2k}^{\star}/ m = \sqrt{s_{2k}} / m$ for $ma = 2$, 6, and 16, which are the three cases we study in detail in the subsequent solutions of the integral equations.

The two-body scattering amplitude has a pole on the real $s_{2k}$ axis, which we call $s_b$. Near the bound state, the scattering amplitude takes the form,
    %%%%%
    \begin{align}
    \lim_{s_{2k}\to s_b}\cM_2(k) =\frac{-g^2}{s_{2k}-s_b},
    \label{eq:M2_pole}
    \end{align}
    %%%%%
where $g$ is the residue at the pole, which can be interpreted as the bound state wave function renormalization factor. Since we are considering the contribution of a pole in the first Riemann sheet below the threshold, the relative momentum of the two-particle subsystem lies on the positive imaginary axis, $q_{2k}=i\kappa_{2k}$, where $\kappa_{2k}>0$ is the binding momentum. In terms of the LO effective range expansion Eq.~\eqref{eq:ERE}, the binding momentum is $\kappa_{2k} = 1/a$. In general, the pole position in $s_{2k}$ can be written in terms of the binding momentum in the standard way,
    %%%%%
    \begin{align}
    s_b = 4(m^2-\kappa_{2k}^2).
    \end{align}
    %%%%%
By equating Eq.~\eqref{eq:ERE} to Eq.~\eqref{eq:M2_pole}, one finds that the residue in the LO in the effective range expansion is given by:
    %%%%%
    \begin{align}
    g=8\sqrt{ 2 \pi \sqrt{s_b}\,\kappa_{2k}  }.
    \end{align}
    %%%%%

    %%%%%%%%%%%%%
    % FIGURE
    %%%%%%%%%%%%%
    %%%%%%%%%%%%%

The integral equations of interest are written in terms of the spectator momentum, therefore we need to define the value of $k$ corresponding to the bound state pole. We will label this ``on-shell" value of $k$ as $q$.  This can be obtained by fixing the two-particle subsystem to be at the bound state pole, and by requiring that systems to be in its total CM frame, the sum of the energy of the two-body subsystem and the spectator satisfies $ E = \sqrt{s_{b}+q^2} + \sqrt{m^2+q^2}$. Solving for $q$ gives:
    %%%%%
    \begin{align}
    \label{eq:qbs}
    q & = \frac{1}{2 E} \, \lambda^{1/2}(E^2,s_b,m^2) \, ,
    \end{align}
    %%%%%
where $\lambda(x,y,z) = x^2 + y^2 + z^2 - 2(xy+yz+zx)$ is the K\"all\'en triangle function. Consequently, the integral equation presented in Eq.~\eqref{eq:d_Sproj} involves an integral over this pole, which makes the numerical convergence of the solutions harder to achieve.

As discussed above in the context of partial wave projection [see Eq.~\eqref{eq:Gs_proj}], it was necessary to introduce a non-zero value of $\epsilon$. Here we once again are required to introduce a non-zero value of $\epsilon$ to mitigate the real-axis pole of $\cM_2$. This shifts the pole slightly away from the axis of integration, allowing for a rigorous definition of the integral. Below the three-particle threshold the exchange propagator never goes on-shell, and thus is a smooth function for all energies in this domain, leaving the two-particle bound state pole as the only singularity in the integration region. Given that it is one of the main issues encountered and addressed in this work, from here on we make the $\epsilon$ dependence explicit in the quantities that are most sensitive to its presence. To that end, we shall denote the $S$ wave projected exchange propagator by $G_{S}(p,k) \to G_{S}(p,k;\epsilon)$, where $G_{S}(p,k;\epsilon)$ is defined exactly as in Eq.~\eqref{eq:Gs_proj}. The $\epsilon$ shift is also implemented in the energy of the two-body system in the following way:
    %%%%%
    \begin{align}
    \label{eq:M2_epsdep}
    \cM_2( k; \epsilon) &\equiv \cM_2(s_{2k} + i\epsilon).
    \end{align}
    %%%%%
This shift propagates through to the amplitudes $\cD$ and $d$, where the latter's integral equation is given by
    %%%%%
    \begin{align}
    \label{eq:duu_epsilon}
    d_S^{(u,u)}(p,k;\epsilon) = -G_S(p,k;\epsilon) - \int_{0}^{\infty} \! \frac{\diff k' \, k'^2}{(2\pi)^2 \omega_{k'}} \, G_{S}(p,k';\epsilon) \, \cM_{2}(k';\epsilon) \, d_S^{(u,u)}(k',k;\epsilon) \, ,
    \end{align}
    %%%%%
where $d_S^{(u,u)}(p,k)$ is given by the limit as $\epsilon \to 0$. After solutions of the integral equation are obtained, it is necessary to analyze their $\epsilon\to 0$ limit numerically. 

Given the definition of $\cD_S$ in Eq.~\eqref{eq:cD_Sproj} one can see that the three-body scattering amplitude has external poles associated with the two-body states. The residue of these is related to the scattering amplitude between the spectator and the two-body bound state. Labeling the spectator as ``$\varphi$" and the bound state as ``$b$", we label this amplitude as $\cM_{\varphi b}$, and denote it as the 2+1 scattering amplitude. More specifically, by continuing the initial and final two-particle subsystems to the bound state poles, the three-body scattering amplitude is related to $\cM_{\varphi b}$ by the LSZ reduction,
    %%%%%
    \begin{align}
    \lim_{s_{2p},s_{2k}\to s_b}
     i{\cM}^{(u,u)}_{3,S}( p, k) &=
    ig \,\frac{i}{s_{2p}- s_b} \,
    i\cM_{\varphi b}( E) \, \frac{i}{s_{2k}- s_b} \, ig .
    \end{align}
    %%%%%
This implies that by evaluating the integral equation for the ${\cM}^{(u,u)}_{3,S}$ as a function of energy one will see this double-pole structure. By determining the residues of these, using the numerical equivalent of 
    %%%%%
    \begin{align}
    \label{eq:phib.continue}
    \cM_{\varphi b}( E) =
    \lim_{s_{2p},s_{2k}\to s_b}
    \frac{(s_{2k}- s_b)\,(s_{2p}- s_b)}{g^2}
    \, {\cM}^{(u,u)}_{3,S}( p,  k),
    \end{align}
    %%%%%
one can determine the $\varphi b\to \varphi b$ scattering amplitude. One important point is the fact that these identities concern the unsymmetrized amplitude. One can obtain the symmetrized amplitude following the procedure defined in Eq.~\eqref{eq:symm_proc}, with eight of the nine terms not contributing to the final sum. This procedure is well defined, however, it requires an additional step in the numerical determination of the solutions of the singular integral equations. Namely, one needs to scan $\cM_{3,S}^{(u,u)}$ as a function of the two-body energy and obtain the residue at the bound state pole. This might be susceptible to numerical instabilities since at this point the zeroes in Eq.~\eqref{eq:phib.continue} need to cancel the divergent pole terms. Fortunately, this numerical issue disappears when performing a redefinition of the three-particle amplitude, as in Eq.~\eqref{eq:d_def} for $\cD$. Such a redefinition allows one to evaluate analytically the cancellation of the external two-body poles and zeros. Given a numerical solution of the integral equations, one would still need to numerically evaluate the limit to the bound state pole, but this would instead be done for a smooth function. Using the expressions above and Eq.~\eqref{eq:d_def}, it is easy to then see that in this limit, $ \cM_{\varphi b}$ can be obtained from $d$ via 
    %%%%%
    \begin{align}
    \lim_{ \cK_\df\to 0} \cM_{\varphi b}( E)
    &=
    g^2 \lim_{s_{2p},s_{2k}\to s_b}
    \, d^{(u,u)}_{S}( p,  k) \, ,
    \end{align}
    %%%%%
The $\varphi b \to \varphi b$ amplitude is our primary subject of study, which, according to the above equation, for chosen $a$ is completely determined by the solution of Eq.~\eqref{eq:d_Sproj}.

By construction, solutions of Eq.~\eqref{eq:d_Sproj} satisfy the three-particle $S$ matrix unitarity~\cite{Briceno:2019muc}. Below the three-particle threshold, the $\varphi b\to\varphi b$ amplitude in turn satisfies the standard $2\to 2$ $S$ matrix unitarity, 
    %%%%%
    \begin{align}
    \label{eq:2p1.unitarity}
    \mathrm{Im} \left[{\cM}^{-1}_{\varphi b} (E)\right] &= -  \, \rho_{\varphi b}(E)  \ ,
    \end{align}
    %%%%%
where 
    %%%%%
    \begin{equation}
    \label{eq:2p1.phase_space}
    \rho_{\varphi b}(E) = \frac{q}{8\pi\,E} \,,
    \end{equation}
    %%%%%
is the two-body phase space between the bound state and the spectator. It follows from Eq.~\eqref{eq:2p1.unitarity} that the amplitude is bounded by unity as $\lvert \rho_{\varphi b}(E) \cM_{\varphi b}(E) \rvert \le 1$ in this kinematic region. Additionally, Eq.~\eqref{eq:2p1.unitarity} tells us that the amplitude can be written using a $2\to 2$ $K$-matrix construction, 
    %%%%%
    \begin{align}
    \label{eq:2p1.Kmat}
    {\cM}_{\varphi b} (E) &= \frac{1}{ \cK_{\varphi b}^{-1}(E) - i  \rho_{\varphi b}(E) } \, ,
    \end{align}
    %%%%%
where $\mathrm{Re} \left[{\cM}^{-1}_{\varphi b} (E)\right] \equiv\cK_{\varphi b}^{-1}(E)$ is real below the $3\varphi$ threshold. The $\varphi b \to \varphi b$ $K$ matrix can be written in terms of a real phase shift $\delta_{\varphi b}$ in the standard way
    %%%%%
    \begin{align}
    \label{eq:2p1.qcot}
    q\cot\delta_{\varphi b}&= 8\pi E \, \cK_{\varphi b}^{-1}(E) \, .
    \end{align} 
    %%%%%
At the $\varphi b$ threshold, $q\cot\delta_{\varphi b}$ reduces to the bound state--spectator scattering length, which we label as $b_0$,
    %%%%%
    \begin{align}
    \label{eq:phib_scatlen}
    \lim_{q\to 0}q\cot\delta_{\varphi b}&=-\frac{1}{b_0}\,.
    \end{align} 
    %%%%%
In the Sec.~\ref{sec:numerics}, we present numerical results for the $\varphi b \to \varphi b$ amplitude below the three-particle threshold. We compare our findings to Ref.~\cite{Romero-Lopez:2019qrt}, which studies the same $\varphi b$ scattering system from the perspective of the associated finite volume formalism~\cite{Hansen:2014eka}.

%%%%%%%%%%%%%%%%%%%%%%%%%%%%%%%%%%%%
%	Section :: Numerical results
%%%%%%%%%%%%%%%%%%%%%%%%%%%%%%%%%%%%
\section{Numerical results}
\label{sec:numerics}

    %%%%%%%%%%%%%
    % FIGURE
    %%%%%%%%%%%%%
    \begin{figure}[t!]
        \centering
        \includegraphics[ width=0.95\textwidth]{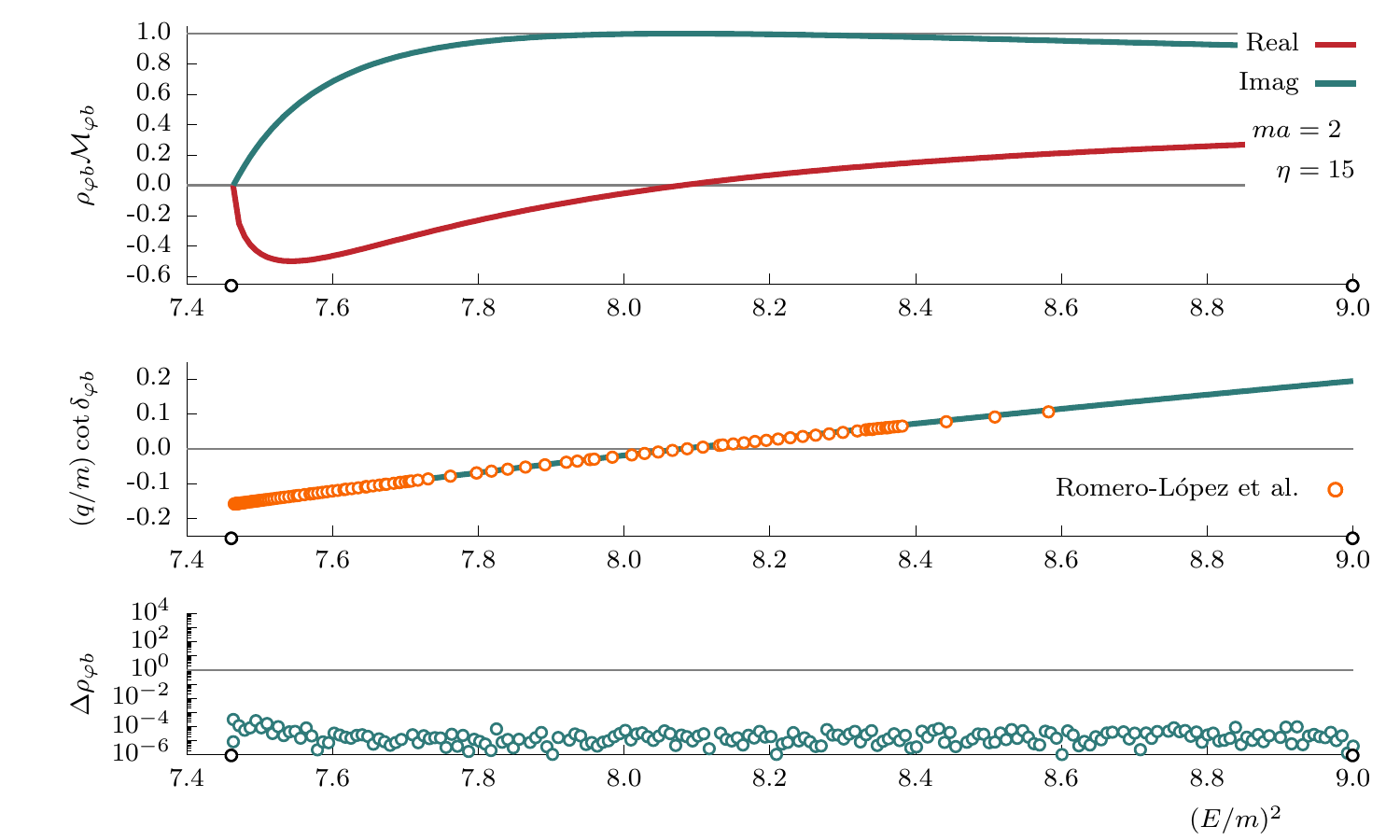}
        \caption{ Solution for the $\varphi b$ scattering amplitude as a function of $(E/m)^2$ below the three-particle threshold for $ma = 2$ obtained using the semi-analytic method solution for $\eta = 15$ as described in the text. The top panel shows the real (red) and imaginary (blue) parts of $\rho_{\varphi b} \cM_{\varphi b}$ where the open circles on the real axis indicate the $\varphi b$ and $3\varphi$ thresholds. The middle panel shows the resulting $q\cot\delta_{\varphi b}$ computed from Eq.~\eqref{eq:2p1.qcot} (blue), with the open orange points being solutions from the three-particle finite volume formalism taken from Ref.~\cite{Romero-Lopez:2019qrt}. The bottom panel shows the unitarity deviation, showing sub-percent level discrepancy as an estimation of the systematic error.
        }
    \label{fig:a2_avg}
    \end{figure}
    %%%%%%%%%%%%%

Having identified the final form of the integral equation we wish to evaluate, namely Eq.~\eqref{eq:d_Sproj} for the partial-wave projected, amputated amplitude $d_S$, we proceed to take the same first steps as in Ref.~\cite{Hansen:2020otl}. The first step amounts to discretize the momenta appearing in the integral equation. This allows one to write the integral equation as a matrix equation that could be solved numerically. We convert the integral to a sum over equidistant mesh points, where the number of points being denoted by $N$. For each value of $E/m$, we compute the solution for various values of $N$, ranging from 1000 to 6000. The solutions of the original integral equations are recovered by extrapolating $N\to\infty$ while keeping $\epsilon N$ fixed. This assures that results are insensitive to both $N$ and $\epsilon$. As one can expect, the results may converge faster or slower depending on the points chosen in the $(N,\epsilon)$ plane. In Sec.~\ref{sec:syst}, we explain the trajectories chosen and the reasoning behind them. We study three cases for the two-particle scattering amplitude, namely $ma = 2, 6, $ and $16$. We first summarize how solutions are computed, and relegate details for the interested reader to the following sections.

Because of the presence of the pole in $\cM_2$, to gain confidence we carry out this procedure in two different ways. In the first method, we follow the procedure outlined in the previous section, where the bound state pole is moved off the real axis. We refer to this as the ``\emph{brute force}" (BF) method. In the second procedure, we evaluate the contribution of the pole analytically. This results in a modified integral equation with a less singular kernel, that we then solve numerically. We refer to this as the ``\emph{semi-analytic}" (SA) method. Detailed descriptions of both of these methods are given in Sec.~\ref{sec:details}.

As discussed in the previous section, solutions must satisfy the $S$ matrix unitarity Eq.~\eqref{eq:2p1.unitarity}.
We use this fact as a check on the quality of solutions as a function of $N$. For each value of $E$, the deviation away from this condition gives us an estimate of the systematic error. In particular, we define
  %%%%%
  \begin{align}
  \label{eq:Deltarho}
  \Delta \rho_{\varphi b} (E;N) \equiv \left| \frac{\mathrm{Im} \left[{\cM}^{-1}_{\varphi b} (E;N)\right] + \, \rho_{\varphi b}(E)}{ \rho_{\varphi b}(E)}\right|\times 100 \ ,
  \end{align}
  %%%%%
which gives a percent measure of the deviation of the solution from unitarity. As $\Delta \rho_{\varphi b} \to 0$, the solution better satisfies the unitarity relation. Therefore, we use this measure to scan for satisfactory solutions and improve the candidate solution by varying $\epsilon$ and $N$ parameters to drive $\Delta \rho_{\varphi b}$ as small as possible. Our goal in this study is to achieve sub-percent level deviation, on the order of $\mathcal{O}(10^{-1}) - \mathcal{O}(10^{-3}) \%$. For most energies, this is easily attainable, with exceptions occurring around points where either $\rho_{\varphi b}$ or the amplitude vanish. 

Since the result of the matrix equation depends on $N$ and $\epsilon$, we must take the ordered double limit as first $N\to \infty$ and then $\epsilon \to 0$. The order of these limits can not be reversed. Qualitatively, as $\epsilon$ steadily decreases the bound state pole moves closer to the axis of integration. This results in the higher and narrower singularity of the integrand in Eq.~\eqref{eq:d_Sproj} and requires greater mesh sizes $N$ to probe the integration kernel effectively. We express $\epsilon$ as a function of $N$ in the form $\epsilon \propto 1/N$, which we derive in Sec.~\ref{sec:Neps_dep}. The proportionality constant is a function of $E$ and includes a controllable parameter which we call $\eta$. We show numerical evidence for this behavior of $\epsilon$ and show that for some restricted values of $\eta$, the resulting solutions are insensitive to our desired working precision, which is a sub-percent level deviation of unitarity. Finally, we perform large $N$ extrapolations to estimate the $N\to\infty$ limit. We find that these extrapolations improve the deviation from unitarity by a few orders of magnitude.

  %%%%%%%%%%%%%
  % FIGURE
  %%%%%%%%%%%%%
  \begin{figure}[t!]
    \centering
    \includegraphics[ width=0.95\textwidth]{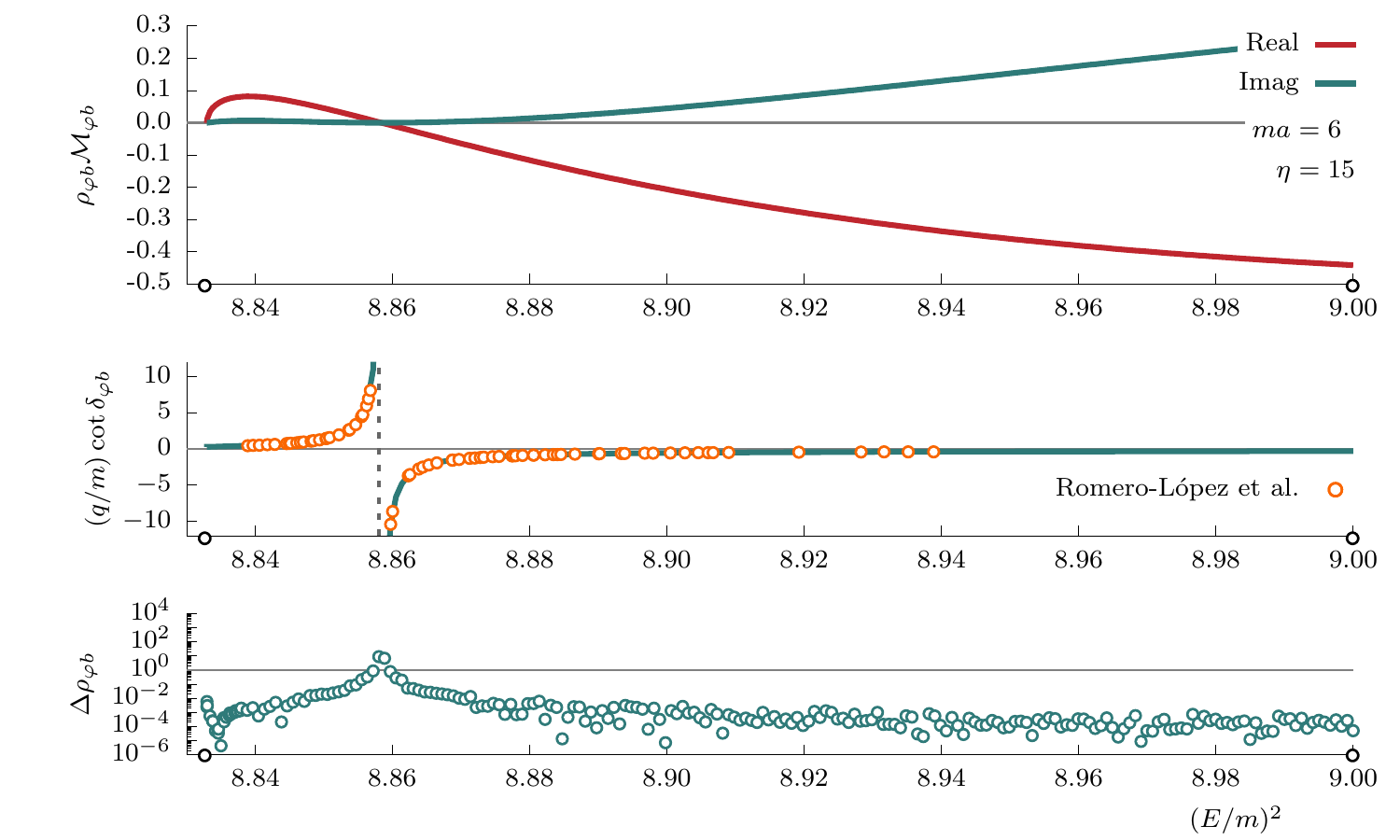}
    \caption{Same as Fig.~\ref{fig:a2_avg} for $ma = 6$.}
  \label{fig:a6_avg}
  \end{figure}
  %%%%%%%%%%%%%

We proceed to the presentation of solutions to the integral equations for total CM energy in the domain $1 + \sqrt{s_b}/m < E/m< 3$. The first case examined is $ma = 2$, which describes a deeply bound state in the two-body subsystem. The resulting bound state plus spectator scattering amplitude is shown in the top panel of Fig.~\ref{fig:a2_avg}. Shown with the vertical dashed line is the $\varphi b$ threshold at $E/m = 1 + \sqrt{3}$. Since below the three-particle threshold, the system is a two-particle system, it is constrained by the usual $S$ matrix principles such as $\lvert \rho_{\varphi b} \cM_{\varphi b} \rvert$ being bounded by unity. The corresponding $q\cot\delta_{\varphi b}$ is shown in the middle panel of Fig.~\ref{fig:a2_avg}, along with points computed of the same system but using the finite volume formalism developed for three-particle scattering processes, Ref.~\cite{Romero-Lopez:2019qrt}. In that work, the authors analytically continue the quantization condition to the $\varphi b$ system. Fitting our solution to an effective range expansion,
  %%%%%
  \begin{equation}
  \label{eq:NumRes.ERE}
  q\cot\delta_{\varphi b} = -\frac{1}{b_0} + \frac{1}{2} r_0 q^2 + \mathcal{O}(q^4) \, ,
  \end{equation}
  %%%%%
where $b_0$ is the $S$-wave scattering length and $r_0$ is the effective range of the bound state plus spectator system, we find the fit values $mb_0 \approx 6.4$ and $mr_0 \approx 2.3$. For these kinematics, we find an excellent agreement with that study. The bottom panel of Fig.~\ref{fig:a2_avg} shows the \textit{unitarity deviation}, which is a measure of the systematic error arising from deviations from the unitarity condition $\mathrm{Im} \, \cM_{\varphi b}^{-1} = -\rho_{\varphi b}$ induced by considering a finite $N$ and $\epsilon$. This $\Delta \rho_{\varphi b}$ function shows sub-percent deviations from this condition and gives a measure of the quality of the solutions for a given energy $E$. For exact solutions, $\Delta \rho_{\varphi b} = 0$. In our calculations for $ma=2$, we find for all energy points except at threshold show sub-percent deviations, indicating that our numerical approximation is satisfactory for the precision we desire in this study. At the threshold, the deviation measures on the order of a percent, which is still exceedingly good for our working precision. Generally, near the threshold we find that solutions show a larger unitarity deviation than other energy points for all $a$, owing to the small kinematic phase space.

In the $ma = 6$ case, the two-body bound state moves toward the threshold, producing a shallow bound state in the two-body subprocess, which results in the $\varphi b\to\varphi b$ amplitude shown in Fig.~\ref{fig:a6_avg}.
Noticeably in this case a zero of the amplitude is slightly below $(E/m)^2 \approx 8.86$. This zero corresponds to a pole in $q\cot\delta_{\varphi b}$ at this kinematic point. Using Eq.~\eqref{eq:NumRes.ERE} to describe this case works in a very limited region close to the threshold due to this pole, which gives a scattering length $m b_0 \approx -3.6$. Deviations from unitarity lie at the sub-percent level except at the threshold and the zero of the amplitude. However, such a large deviation from unitarity is not a major concern since it results in a percent-level systematic error for an observable which is equal to zero at this point.

\begin{figure}[t!]
        \centering
        \includegraphics[ width=0.95\textwidth]{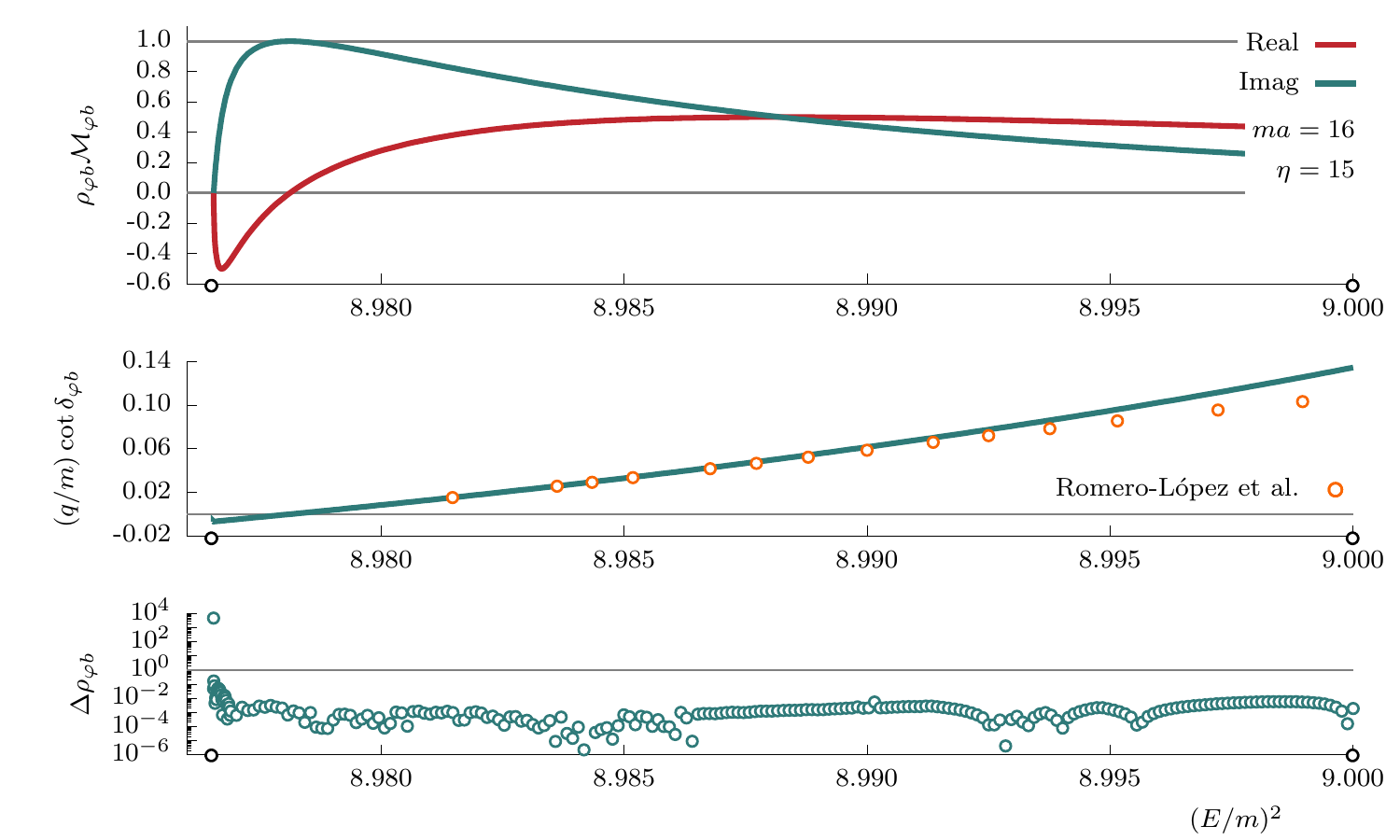}
        \caption{Same as Fig.~\ref{fig:a2_avg} for $ma = 16$.}
    \label{fig:a16_avg}
    \end{figure}

Finally, for $ma = 16$ there is a very shallow bound state in the two-body channel. In Fig.~\ref{fig:a16_avg}, the top, middle, and bottom panels show the resulting amplitude, $q\cot\delta_{\varphi b}$, and $\Delta \rho_{\varphi b}$, respectively. We find the corresponding $mb_0 \approx 150$, consistent with Ref.~\cite{Romero-Lopez:2019qrt}. However, we observe a significant deviation of the $q\cot\delta_{\varphi b}$ as compared to Ref.~\cite{Romero-Lopez:2019qrt} for energies near the three-body threshold. These are attributed to the failure of the method used in Ref.~\cite{Romero-Lopez:2019qrt} near and above the three-particle region.

Given this slight deviation, it is worthwhile to summarize the method used in Ref.~\cite{Romero-Lopez:2019qrt} and explain its errors in this kinematic region. There, finite-volume energies levels for these toy theories using large volumes, $mL=20-70$, were obtained. For these volumes, some states lie below the $3\varphi $ threshold and above the $\varphi b$ threshold. If these are sufficiently below the $3\varphi $, these can be approximated as two-body finite-volume states and must therefore satisfy the two-body quantization condition~\cite{Luscher:1986pf,Luscher:1990ux}, from which one can obtain $q\cot\delta_{\varphi b}$. As the finite-volume approaches the $3\varphi $ threshold, it no longer satisfies the two-body quantization condition. In general, this transition is not an abrupt, but rather a continuous behavior. This is consistent with the deviation between the results found here, which are not susceptible to these issues, and that obtained ibid.

    %%%%%%%%%%%%%
    % FIGURE
    %%%%%%%%%%%%%
    %%%%%%%%%%%%%

\begin{figure}[t!]
\centering
\includegraphics[ width=0.95\textwidth]{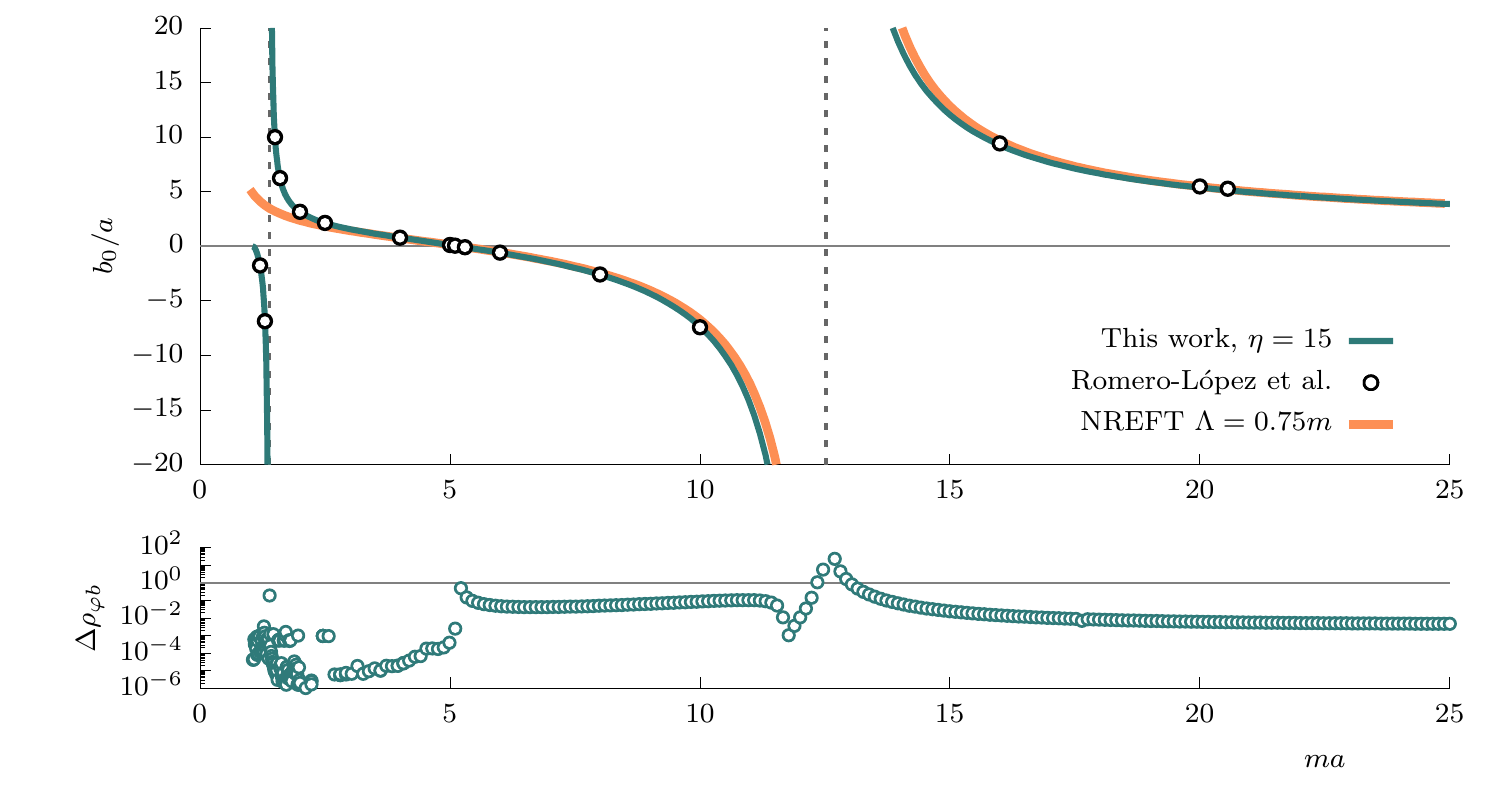}
\caption{Ratio of the $\varphi b$ scattering length $b_0$ to the two-particle scattering length $a$ as a function of $ma$. Our solution (blue) is computed using the semi-analytic method with $\eta = 15$ as described in the text. The lower panel shows the unitarity deviation for the solution, showing sub-percent deviation as an estimation of systematic error for most $ma$. Vertical dashed lines show the locations of the asymptotes. Open black circles are solutions computed using the three-particle finite volume formalism as taken from Ref.~\cite{Romero-Lopez:2019qrt}. The orange line shows the result as computed with an NREFT formalism as described in Ref.~\cite{Bedaque:1998km} using a cutoff $\Lambda = 0.75 m$.
}
\label{fig:bva}
\end{figure}

As a final comparison to Ref.~\cite{Romero-Lopez:2019qrt}, we solve the amplitude at threshold energy $E = m + \sqrt{s_b}$, and compute the ratio of the $\varphi b$ scattering length to the two-particle scattering length $b_0 / a$ as a function of $ma$. This result is shown in Fig.~\ref{fig:bva}, where the blue line is the result of our calculation using the extrapolation method as described above, and the open black circles are the results from Ref.~\cite{Romero-Lopez:2019qrt}, which uses the finite volume formalism to extract the $\varphi b$ phase shift. We also compare our result to one computed from a non-relativistic effective field theory (NREFT) formalism~\cite{Bedaque:1998km}, where we follow Ref.~\cite{Romero-Lopez:2019qrt} by choosing a cutoff $\Lambda = 0.75 m$.cWe find that for small $ma$, corresponding to highly relativistic systems, the solution differs from the NREFT considerably with the presence of an additional pole as compared to the NREFT result.  Near the pole at $ma \approx 13$, our solution begins to deviate from the NREFT solution.  Two factors can attribute to such a discrepancy, one of which is the fact that the NREFT result is computed at a finite matrix size of $N=2000$ whereas we perform an extrapolation on our solutions. The second is that the solution is scheme dependent through the parameter $\Lambda$, where we chose the value shown in order to compare to the result shown in Ref.~\cite{Romero-Lopez:2019qrt}.

%%%%%%%%%%%%%
% FIGURE
%%%%%%%%%%%%%
%%%%%%%%%%%%%

%%%%%%%%%%%%%%%%%%%%%%%%%%%%%%%%%%%%
%	Section :: Details of numerical methods
%%%%%%%%%%%%%%%%%%%%%%%%%%%%%%%%%%%%
\section{Details of numerical methods }
\label{sec:details}

In this section, we discuss the procedure of converting the integral equations to matrix equations and define both the BF and SA methods introduced in the previous section. To evaluate the integral appearing in Eq.~\eqref{eq:d_Sproj}, it is necessary to discretize the spectator momenta in either method. We denote these momenta by a discrete index, i.e. we make the replacement $k'\to k'_n$, and generate a uniform mesh of points. The minimum value that the momenta can take is $k_{\text{min}} = 0$, while the maximum value corresponds to the point at which the cutoff function $H$, defined in Eq.~\eqref{eq:cutoff}, is zero. It vanishes when $E^{\star}_{2,k}=0$, which implies $k_{\rm max}^2 = \left((E^{2}-m^2)/2E\right)^2$. We replace the measure $dk' \to \Delta k' =k'_{\mathrm{max}} (E) / N$ which is the distance between mesh points for a given energy. We find that because the integrand is singular, it is necessary to use a large number of mesh points to finely sample the kernel.\footnote{Alternatively, one can implement an improved meshing technique, which probes more densely the regions in momentum $k'$ where the integral equation kernel is varying rapidly. In our case, such meshing techniques would interpolate the vicinity of the pole more accurately and in consequence, could require smaller mesh sizes. However, such improvement schemes are subject to an increased number of tunable parameters compared to the uniform meshing, producing additional systematic to our results. In the course of our work, we have tried three different improved meshing procedures, which are not described here for the simplicity of the presentation and to avoid these additional systematic effects.}

%%%%%%%%%%%%%%%%%%%%%%%%%%%%%%%%%%%%
%	Section :: Brute force method
%%%%%%%%%%%%%%%%%%%%%%%%%%%%%%%%%%%%
\subsection{Brute force method}

Having a uniform mesh of points, one proceeds to discretize the momentum appearing in Eq.~\eqref{eq:d_Sproj}.
In practice, to solve this equation it is necessary to keep $N$ fixed as a finite parameter and test the convergence with $N$ and $\epsilon$.  Making the $N$ dependence of $d$ explicit we have
    %%%%%
    \begin{align}
    \label{eq:dNeps_eq}
    d^{(u,u)}_{S}(  p,   k;\epsilon, N) 
    &=
    - G_{S}( p,  k;\epsilon) 
    - \sum^{N-1}_{n=0} \frac{\Delta k' \, k'^2_n }{ (2\pi)^2 \, \omega_{k'_n}}
    \, G_{S}(  p,   k'_n;\epsilon)
    \, \cM_2( k'_n ;\epsilon) \, 
    d^{(u,u)}_{S}( k'_n,   k;\epsilon, N) \, .
    \end{align}
    %%%%%

To solve this we write $d$ as a matrix in the space defined by the set of $\{k_n'\}$, with matrix elements
    %%%%%
    \begin{align}
    d^{(u,u)}_{S;  {n}{n'}} 
    &=d^{(u,u)}_{S}(  k_{n},   k_{n'} ;\epsilon, N) \, .
    \end{align}
    %%%%%
This allows us to write Eq.~\eqref{eq:dNeps_eq} as a simple linear system, with a solution 
    %%%%%
    \begin{align}
    \label{eq:Mphib_sol}
    d^{(u,u)}_{S} (  p  , k; \epsilon, N)
    &=
    - \left[ B^{-1} \, {G}_{S}\right]_{nn'}\bigg|_{k_n=p,\,k_{n'}=k} \ ,
    \end{align}
    %%%%%
where $B$ is a matrix defined as
    %%%%%
    \begin{align}
    \label{eq:Bphib}
    B_{n {n'}}
    &=
    \delta_{k_{n},k_{n'}}
    + \frac{ \Delta k' \, k_{n'}^2 }{ (2\pi)^2 \, \omega_{k_{n'}}}
    \, G_{S}(k_{n}, k_{n'}; \epsilon)
    \, \cM_2( k_{n'}; \epsilon)  \, .
    \end{align}
    %%%%%

Note, although above we assumed $d$ is a matrix, it is sufficient to assume only $p\in \{ k_{n}\}$ and leave $k$ as a continuous variable. Since we are interested in the $\varphi b \to \varphi b$ amplitude, we choose $k = q$.
For a large value of $N$, using Eq.~\eqref{eq:dNeps_eq}, one easily interpolates to any continuous value of these momenta within the kinematically allowed values, including the on-shell point $q$ for bound state plus spectator system, defined in Eq.~\eqref{eq:qbs}.

In Sec.~\ref{sec:syst} we explain how one may assess systematic errors of solution to Eq.~\eqref{eq:dNeps_eq}. In particular, to arrive at the solution to the integral equation one must take the ordered-double limit, 
    %%%%%
    \begin{align}
    \label{eq:double_limit}
    d^{(u,u)}_{S} (  p  , k) = \lim_{\epsilon \to 0} \lim_{N\to\infty}d^{(u,u)}_{S} (  p  , k ; \epsilon, N). 
    \end{align}
    %%%%%
In practice, by calculating the amplitude for several values of sufficiently large $N$ and small enough $\epsilon$ one could perform a careful extrapolation of the numerical result and test the convergence. To make the extrapolation of the convergence tests systematic, in Sec.~\ref{sec:Neps_dep} we determine the asymptotic behavior of the error for large $N$ and small $\epsilon$, finding
    %%%%%
    \begin{align}
    d^{(u,u)}_{S} ( p, k) = 
    d^{(u,u)}_{S} ( p, k; \epsilon, N)
    + \mathcal{O}\left(e^{-\eta}\right),
    \end{align}
    %%%%%
where $\eta \equiv 2\pi N\epsilon_q / k_{\mathrm{max}}$ and $\epsilon_q$ is function of the energy and is linearly proportional to $\epsilon$ defined in Eq.~\eqref{eq:epsq_def}. This $\eta$ parameter provides a relation between $\epsilon$ and $N$, which for a fixed $\eta$ the ordered double limit is ensured. This explains explicitly that for a given value of $N$ one cannot make $\epsilon$ arbitrarily small, otherwise, the error introduced will no longer be exponentially suppressed. A sufficient condition is that for a given $\epsilon_q$ the matrix size $N$ must be large enough, satisfying inequality $\eta \gg 1$.  In Sec.~\ref{sec:sys.eta} we provide numerical evidence of this behavior of the error.

%%%%%%%%%%%%%%%%%%%%%%%%%%%%%%%%%%%%
%	Section :: Semi-analytic method
%%%%%%%%%%%%%%%%%%%%%%%%%%%%%%%%%%%%
\subsection{Semi-analytic method}

Here we consider an alternative method of solution where we evaluate analytically the contribution due to the $\cM_2$ pole. To do this we add and subtract the pole contribution to $\cM_2$, given in Eq.~\eqref{eq:M2_pole}. In particular, we aim to isolate the $\delta$-function contribution to the pole arising from the imaginary part. We define a pole-subtracted two-body scattering amplitude $\Delta \cM_2$ by
    %%%%%
    \begin{align}
    \label{eq:delta-m-2}
    \Delta \cM_{2}(s_{2k'};\epsilon)
    &\equiv
    \cM_2(s_{sk'};\epsilon)
    - g^2\, i\pi \delta_\epsilon (s_{2k'}-s_b) \, ,
    \end{align}
    %%%%%
where we introduced an $\epsilon$-dependent delta function, which we denote $\delta_{\epsilon}$.
It is defined as
    %%%%%
    \begin{align}
    \delta_\epsilon (s_{2k'}-s_b)&=\frac{\epsilon}{\pi \left((s_{2k'}-s_b)^2 + \epsilon^2\right)},
    \end{align}
    %%%%%
and reproduces the Dirac delta function for $\epsilon \to 0$. It is straightforward to check that with this definition of $\delta_\epsilon$, $\Delta \cM_{2,\epsilon}$ is finite at the bound state pole and it contains the same branch cut as $\cM_2$. The $\epsilon$-regulated delta is necessary since $\cM_2$ only supports the Dirac delta function form for the imaginary part at the bound state pole.

With the definition of Eq.~\eqref{eq:delta-m-2}, we can isolate the pole in $\cM_2$ by writing it as
    %%%%%
    \begin{align}
    \cM_2(s_{2k'}) 
    &= g^2\, i\pi \delta(s_{2k'} - s_b) + 
    \lim_{\epsilon\to 0}\Delta \cM_{2}(s_{2k'};\epsilon).
    \end{align}
    %%%%%
Given that the pole is evaluated in terms of $s_{2k}$, we perform the change of variables in the integral with the pole,
    %%%%%
    \begin{align}
    \label{eq:SA.jacobian}
    \int_{0}^{k_{\mathrm{max}}} \! \! \frac{\mathrm{d}k \, k^2}{(2 \pi)^2\, \omega_{k}}
    =
    \int_{0}^{(E-m)^2} \!\! \frac{ \mathrm{d} s_{2k'} }{16\pi^2 E^2} \,  \lambda^{1/2}(E^2, s_{2k'}, m^2) \, .
    \end{align}
    %%%%%
With these identities, we proceed to modify the integral equation for $d$, Eq.~\eqref{eq:d_Sproj}, arriving to
    %%%%%
    \begin{align}
    \label{eq:Mphib_eqv2}
    d^{(u,u)}_{S}(  p,   k; \epsilon) 
    &=
    - {G}_{S}( p,  k;\epsilon) - i g^2 \, {G}_{S}(  p,   q;\epsilon) \, \rho_{\varphi b} (E) \, d^{(u,u)}_{S}( q,  k; \epsilon) \nn\\[5pt] & \hspace{3cm} -\int \! \! \frac{\mathrm{d} k' \, k'^{2}}{(2 \pi)^2\, \omega_{k'}} \, {G}_{S}(  p,   k';\epsilon) \, { \Delta  {\cM}_2( k' ;\epsilon)} \, d^{(u,u)}_{S}( k',  k; \epsilon) \, ,
    \end{align}
    %%%%%
where in the last line we have used Eq,~\eqref{eq:qbs} and \eqref{eq:2p1.phase_space} after we integrated over the delta function with Eq.~\eqref{eq:SA.jacobian}. In the resulting integral equation, the remaining integral has at worst logarithmic singularities due to the exchange propagator.

Just as for the BF method, we introduce a uniform mesh to write this as,
    %%%%%
    \begin{align}
    \label{eq:d_semi_Neps_eq}
    d^{(u,u)}_{S}(  p,   k;\epsilon, N) 
    &=
    - {G}_{S}( p,  k;\epsilon) 
    - i g^2 \, {G}_{S}(  p,   q;\epsilon)
    \, \rho_{\varphi b} (E)
    \, d^{(u,u)}_{S}( q,  k;\epsilon, N)
    \nn\\[5pt]
    & \hspace{3cm}
    - \sum^{N-1}_{n=0} \frac{ \Delta k' \,
    k_{n'}^2}{ (2\pi)^2 \, \omega_{k_{n'}}}
    \, G_{S}(  p, k'_n;\epsilon)
    \Delta \cM_2( k'_n ;\epsilon)
    \, d^{(u,u)}_{S}( k'_n, k;\epsilon, N) \, .
    \end{align}
    %%%%%
This can then be solved in general by following a two-step process. First, solve for $d_S^{(u,u)}$ when $p=q$. Having this, one then readily insert this back into the second term of the right-hand side of the equation and solve for $d_S^{(u,u)}$ for arbitrary values of $p$. We proceed by introducing 
    %%%%%
    \begin{align}
    \Delta B_{nn'}
    &=
    \delta_{k_{n}, k_{n'}}
    + \frac{ \Delta k' \,
    k_{n'}^2}{ (2\pi)^2 \, \omega_{k_{n'}}}  \, G_{S}(k_{n}, k_{n'}; \epsilon)
    \, \Delta \cM_2( k_{n'}; \epsilon) \, .
    \label{eq:dB}
    \end{align}
    %%%%%
As before, this allows us to write Eq.~\eqref{eq:d_semi_Neps_eq} as a matrix equation ,
    %%%%%
    \begin{align}
    \label{eq:d_semi_mat}
    \left[\Delta B\,
    d^{(u,u)}_{S}\right]_{   nn'} 
    &=
    - {G}_{S, nn'}
    - i g^2 {G}_{S}(  k_n,   q;\epsilon)
    \, \rho_{\varphi b} (E) \,
    d^{(u,u)}_{S}( q,  k_{n'};\epsilon, N).
    \end{align}
    %%%%%
By multiplying both sides by the inverse of $\Delta B$, proceeding to set $k_n=q$, we arrive at an algebraic equation for $d^{(u,u)}_{S}( q,  k_{n'},\epsilon, N)$, whose solution is
    %%%%%
    \begin{align}
    \label{eq:eq:d_semi_qsol}
    g^2 d^{(u,u)}_{S}( q,  k_{n'};\epsilon, N)
    &=
    \frac{ \left[{\cK}_{\varphi b}\right]_{nn'}
    }{ 1- i  \left[ {\cK}_{\varphi b} \,  \rho_{\varphi b} \right]_{nn''} }
    \bigg|_{k_n=k_{n''}=q} \,
    \end{align}
    %%%%%
where 
    %%%%%
    \begin{align}
    {\cK}_{\varphi b,   nn'} 
    \equiv - g^2 \left[\Delta B^{-1} {G}_{S} \right]_{nn'},
    \end{align}
    %%%%%
is an off-shell extension of the $K$ matrix of the $\varphi b$ system. In other words, when $k_n,k_{n'}=q$ this coincides with the physical $\varphi b$ $K$ matrix as defined in Eq.~\eqref{eq:2p1.Kmat}. In principle, $\cK_{\varphi b}$ must be a real function below the three-particle threshold. However, since we work with finite $N$, $\cK_{\varphi b}$ may be complex. The level of complexity is measured by the same $\Delta\rho_{\varphi b}$ as defined in Eq.~\eqref{eq:Deltarho}. As in the case with the BF method, one must then take the ordered, double limit defined Eq.~\eqref{eq:double_limit}. 

%%%%%%%%%%%%%%%%%%%%%%%%%%%%%%%%%%%%
%	Section :: Comparing BF and SA
%%%%%%%%%%%%%%%%%%%%%%%%%%%%%%%%%%%%
%\subsection{Comparing brute force and semi-analytic methods}

In Sec.~\ref{sec:numerics} we showed results for the $\varphi b$ amplitude as computed using the SA method. Both the BF and SA methods reliably recover the amplitude at the various $N$ we tested which were consistent with each other. We found that the SA method for most kinematics can yield results that have a unitarity deviation an order of magnitude lower than the BF method. Since the results at the precision at which we show plots are visually indistinguishable, we show only the results for the SA method throughout this article.

%%%%%%%%%%%%%%%%%%%%%%%%%%%%%%%%%%%%
%	Section :: Assessing systematics
%%%%%%%%%%%%%%%%%%%%%%%%%%%%%%%%%%%%
\section{Assessing systematics}
\label{sec:syst}

Since the integral equations in both methods are always solved numerically at a finite $N$, and for a nonzero $\epsilon$, our solutions deviate systematically from the result defined by the $N\to \infty$ and $\epsilon \to 0$ limits. In this section, we present a detailed discussion of these systematic effects.

%%%%%%%%%%%%%%%%%%%%%%%%%%%%%%%%%%%%
%	Section :: Proof of systematic error
%%%%%%%%%%%%%%%%%%%%%%%%%%%%%%%%%%%%
\subsection{Proof of the $\mathcal{O}\left(e^{-\eta }\right)$ systematic error}
\label{sec:Neps_dep}

Our first step is to evaluate the difference between the solution of the desired integral equation, given in Eq.~\eqref{eq:d_Sproj}, and the numerical solution, which satisfies Eq.~\eqref{eq:dNeps_eq}, and is obtained for a finite value of $N$. Both the integral equation and the  matrix equation require the introduction of an $\epsilon$, which must be set to zero at the end of the computation. Here, we consider the difference of these at the stage where $\epsilon$ is nonzero. Assuming our isotropic mesh, we find that it is given by
    %%%%%
    \begin{align}
    \sigma_{\varphi b}(  p,   k; \epsilon , N)
    & \equiv \left| d^{(u,u)}_{S}(p, k; \epsilon) -  d^{(u,u)}_{S}(p, k; \epsilon, N) \right|
    \nn\\
    %%%
    & \approx
    \left| \left[ \int_0^\infty \mathrm{d}k'
    - \frac{k_{\mathrm{max}} }{N}
    \sum^{N-1}_{n=0} \,\right]
    \, G_{S}(p, k'; \epsilon)
    \, k'^2 \frac{ \cM_2( k'; \epsilon)}{ (2\pi)^2 \, \omega_{k'}} \, d^{(u,u)}_{S}(k', k; \epsilon) \right|
    \nn\\
    %%%
    &=
    \left| \sum^{N-1}_{n\neq 0}
    \, \int_0^\infty \mathrm{d}k'
    \, \frac{k'^2\, e^{ i 2\pi\,nk'N/k_{\mathrm{max}} } }{{ (2\pi)^2 \, \omega_{k'}}}
    \ \, G_{S}(p, k'; \epsilon) { \cM_2( k'; \epsilon)}
    \, d^{(u,u)}_{S}(k', k;\epsilon) \right|.
    \label{eq:sigma_phib}
    \end{align}
    %%%%%
In the second line, we used $d^{(u,u)}_{S}(p, k; \epsilon, N) = d^{(u,u)}_{S}(p, k; \epsilon) + \mathcal{O}(\sigma(p,k;\epsilon,N)) $ and kept only the leading term. This difference is then written as a combined summation and integration using the Poisson summation formula in the third line. In general, the integral is saturated by its singularities. In this case, we have the pole singularity due to $\cM_2$, as well as logarithmic ones. Given that the former is the largest, and consequently the source of the leading error, we approximate the integral by the contribution due to the pole which is at
    %%%%%
    \begin{align}
    \label{eq:epsq_def}
    k &= \frac{\lambda(s_b - i \epsilon, m^2, E^2)}{2 E} \nn \\
    %%%
    &\approx
    q + i\epsilon \, \left( \frac{E^2 + m^2 - s_b}{4 q E^2 } \right)
    \equiv q+i\epsilon_q. 
    \end{align}
    %%%%%
Above the $\varphi b$ threshold, $\epsilon_q(E)>0$, as expected. With this, we obtain the correction near the pole, which is
    %%%%%
    \begin{align}
    \label{eq:sigma_phibf}
    \sigma_{\varphi b}(p, k; \epsilon , N)
    &\approx
    \left| \sum^{N-1}_{n\neq 0}
    \, e^{ i 2\pi\,n q N / k_{\mathrm{max}} } 
    \, e^{ - 2\pi\,n \epsilon_q N / k_{\mathrm{max}}}
    \, G_{S}(p, q;\epsilon) \rho_{\varphi b} (E)
    \, d^{(u,u)}_{S}(q, k;\epsilon) \right|
    \nn\\
    %%%
    & \approx
    \left|
    \, G_{S}(p, q;\epsilon) \rho_{\varphi b} (E)
    \, d^{(u,u)}_{S}(q, k;\epsilon)
    \right| e^{ -\eta},
    \end{align}
    %%%%%
where in the last equality we have defined $\eta \equiv 2\pi  \epsilon_q N / k_{\mathrm{max}} $ and assumed $\eta >1$ while ignoring contributions from $n>1$ that are further suppressed. This tells us the condition needed for the systematic error to be suppressed, 
    %%%%%
    \begin{align}
    \label{eq:eps_bound}
    \eta \gg 1 \, .
    \end{align} 
    %%%%%
The parameter $\eta$ characterizes dependence of $\epsilon$ as a function of $N$. For a fixed $\eta$, the ordered double limit in $\epsilon$ and $N$ is ensured by a single limit in $N$. Our derivation does not completely fix the $N$ dependence of the function, as we have only looked at the dominating term in the series of the NLO correction. Moreover, since the $N$-dependent $\epsilon$ is propagated through to the amplitude in a form of an energy shift, this gives an explicit $N$ dependence to the energy. Moreover, although the above bound is derived for the explicit pole in $\cM_2$, and thus is applicable to the BF method, we use the same constant $\eta$ trajectory in the SA method in order to ensure the double limit is properly taken.

%%%%%%%%%%%%%%%%%%%%%%%%%%%%%%%%%%%%
%	Section :: Large N extrapolations
%%%%%%%%%%%%%%%%%%%%%%%%%%%%%%%%%%%%
\subsection{Large $N$ extrapolations}

    %%%%%%%%%%%%%
    % FIGURE
    %%%%%%%%%%%%%
    \begin{figure}[t!]
        \centering
        \includegraphics[ width=0.49\textwidth]{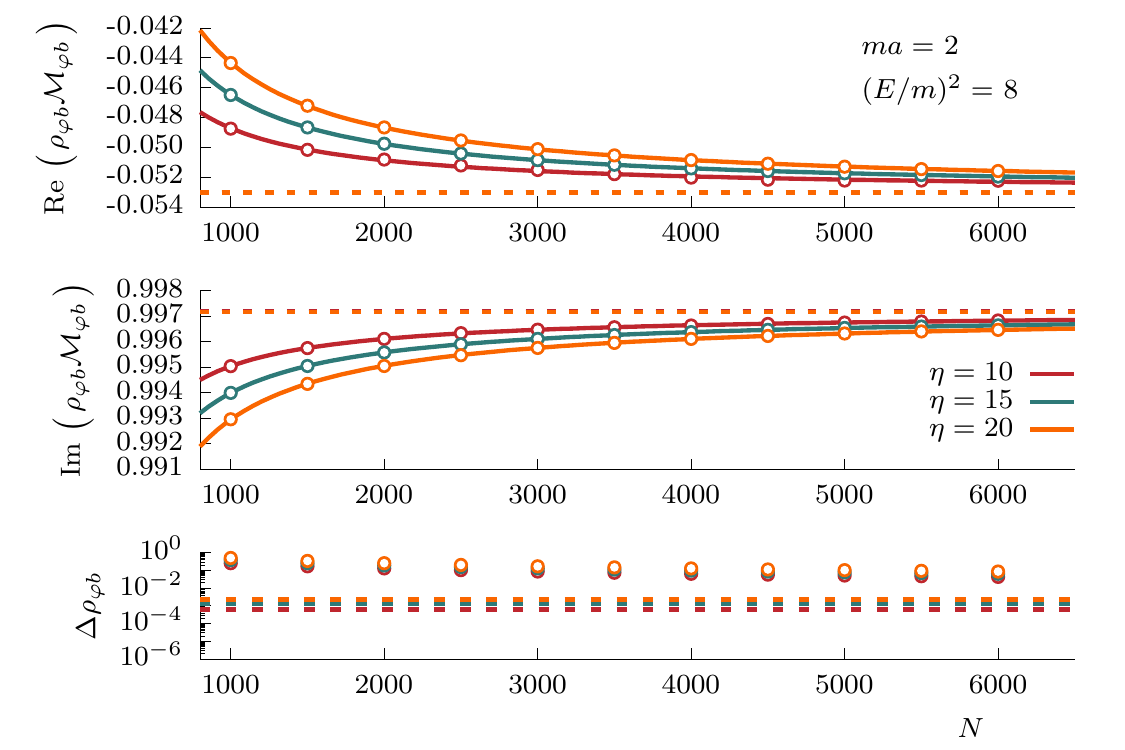}
        \includegraphics[ width=0.49\textwidth]{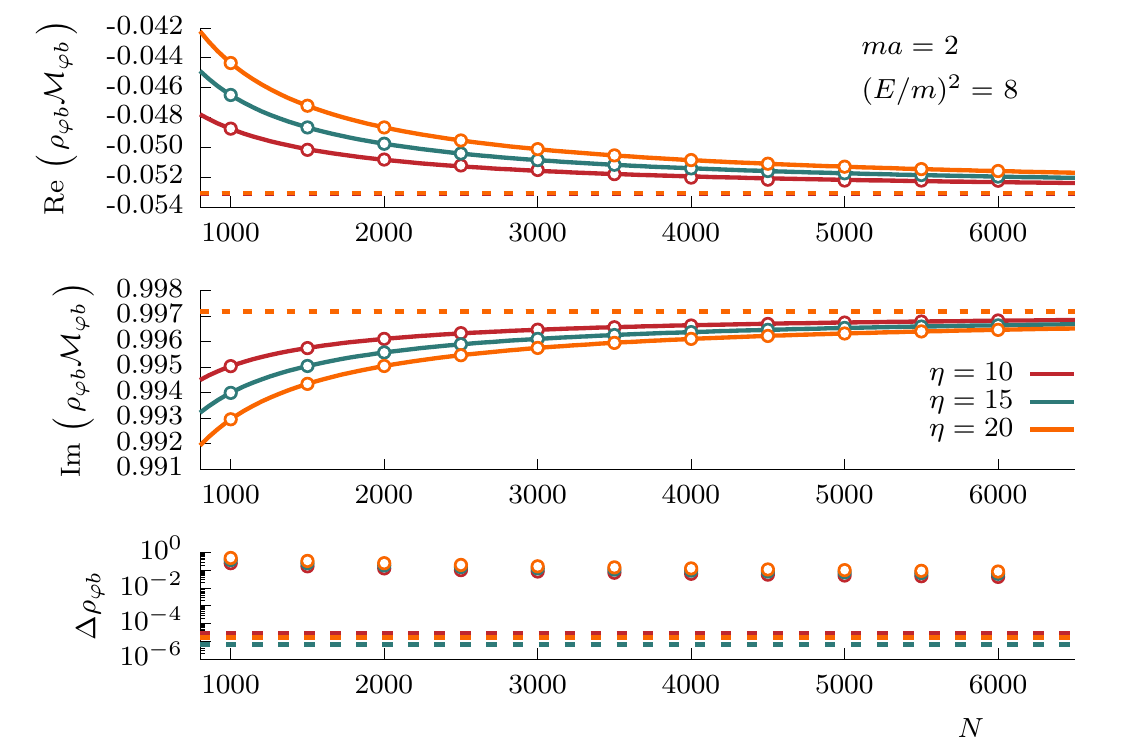}
	\put(-477,2){(a)}
	\put(-225,2){(b)}
        \caption{Example of a large $N$ extrapolation for $ma = 2$ and $(E / m)^2 = 8$ for (a) linear and (b) quadratic fit models. Top panel shows data and fit for the real part of $\rho_{\varphi b} \cM_{\varphi b}$, middle panel shows the imaginary part, and the bottom panel shows the unitarity deviation. Three different data sets and fits are included for $\eta = 10$ (red), $\eta = 15$ (blue), and $\eta = 20$ (orange) parameters. }
    \label{fig:a2_eta15_Nextra}
    \end{figure}
    %%%%%%%%%%%%%

Our solutions are obtained from the matrix equations, Eqs.~\eqref{eq:dNeps_eq} and~\eqref{eq:d_semi_Neps_eq}, therefore they explicitly depend on the mesh size $N$. Moreover, from the previous section, we assert that $\epsilon \propto 1/N$, which introduces additional $N$-dependencies in both the BF and SA methods. As discussed above, fixing $\eta \gg1$ assures that the leading $N$ behavior is due to the shift in the energy rather than the discretization. This tells us that at large $N$, the finite $N$ amplitude for a given $E/m$, $ma$, and $\eta$ can be represented as the $N\to\infty$ amplitude and a correction factor in $1/N$,
    %%%%%
    \begin{align}
    \cM_{\varphi b}(E;N) = \cM_{\varphi b}(E) + \mathcal{O}(1/N) \, .
    \end{align}
    %%%%%
Therefore, as $N\to\infty$ the numerical solution better approximates the continuum solution for $\cM_{\varphi b}$. Since we always work with a finite $N$, we employ a program which solves the integral equation for various values of $N$, and perform an extrapolation in $N$ in order to estimate the $N\to\infty$ amplitude. The extrapolation is executed by fitting the finite $N$ amplitude for a fixed $E/m$, $ma$, and $\eta$ to a function which has a polynomial dependence on $1/N$. We choose two kinds of fit models: linear and quadratic in $1/N$,
    %%%%%
    \begin{subequations}
    \label{eq:polynomial}
    \begin{alignat}{2}
    \cM_{\varphi b}(E;N) & = \cM_{\varphi b}(E) + \frac{\alpha}{N} &&\qquad (\mathrm{linear})  \, , \\
    %%%
    & = \cM_{\varphi b}(E) + \frac{\alpha}{N} + \frac{\beta}{N^2} &&\qquad (\mathrm{quadratic}) \,,
    \end{alignat}
    \end{subequations}
    %%%%%
where $\alpha$ and $\beta$ are complex parameters for each energy. Although the extrapolated amplitude is our best estimate of the $N \to \infty$ solution, since our calculations are always performed at finite $N$, there is a systematic error that propagates through to the extrapolated value. This impacts the result, which can be measured e.g. through a $\Delta \rho_{\varphi b}$ test computed with the extrapolated value of the amplitude. In Fig.~\ref{fig:a2_eta15_Nextra} we show examples of extrapolations for both the linear and quadratic models for $ma = 2$, $(E/m)^2 = 8$, and $\eta = 10$, 15, 20. The extrapolations produce improvements on the asymptotic deviation of unitarity by a few orders of magnitude as compared with the finite $N$ values. We see evidence that the $\eta$-dependence of the solutions are mild when compared to the finite $N$ effects.

To obtain the final amplitudes presented in Sec.~\ref{sec:numerics}, we generate a small ensemble of extrapolated solutions by fitting both linear and quadratic models to various subsets of our finite $N$ results. In our computations, we used 11 equidistant points in the interval $1000 \leq N \leq 6000$. For both the linear and quadratic models, we repeat the fit by successively excluding the lowest value of $N$ until we reach $N=4000$. This results in a total of 14 fits, from which we choose the one which minimizes the extrapolated $\Delta \rho_{\varphi b}$. To estimate a systematic error associated with the $N$-dependence, we take the difference between this fit and the one which maximizes $\Delta \rho_{\varphi b}$ as twice our systematic error associated with $N$. The impact of this error is orders of magnitude smaller than the results themselves and thus invisible in the presented results. We find, as shown in the main results, that our estimated unitarity deviation is sub-percent level for all kinematic points except for those where the amplitude is zero, and thus Eq.~\eqref{eq:Deltarho} becomes numerically ill-defined.

%%%%%%%%%%%%%%%%%%%%%%%%%%%%%%%%%%%%
%	Section :: eta independence of solutions
%%%%%%%%%%%%%%%%%%%%%%%%%%%%%%%%%%%%
\subsection{$\eta$ independence of solutions}
\label{sec:sys.eta}

\begin{figure}[t!]
\centering
\includegraphics[ width=0.5\textwidth]{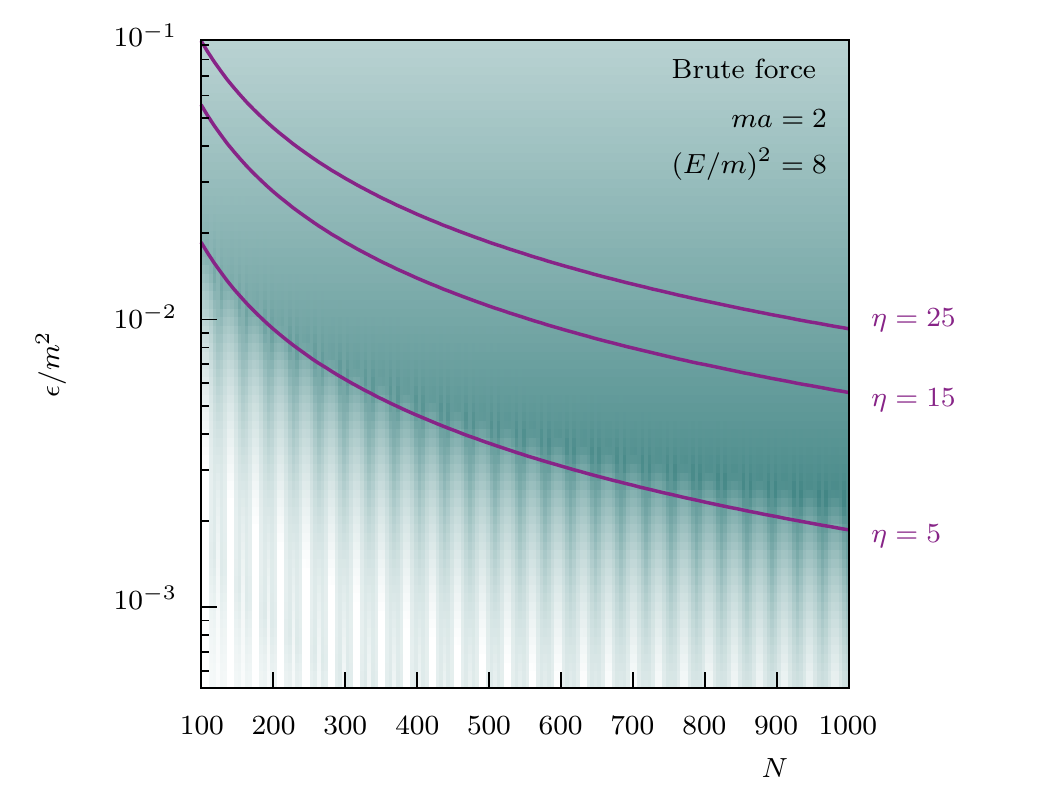}\hspace{-1.2cm}
\includegraphics[ width=0.5\textwidth]{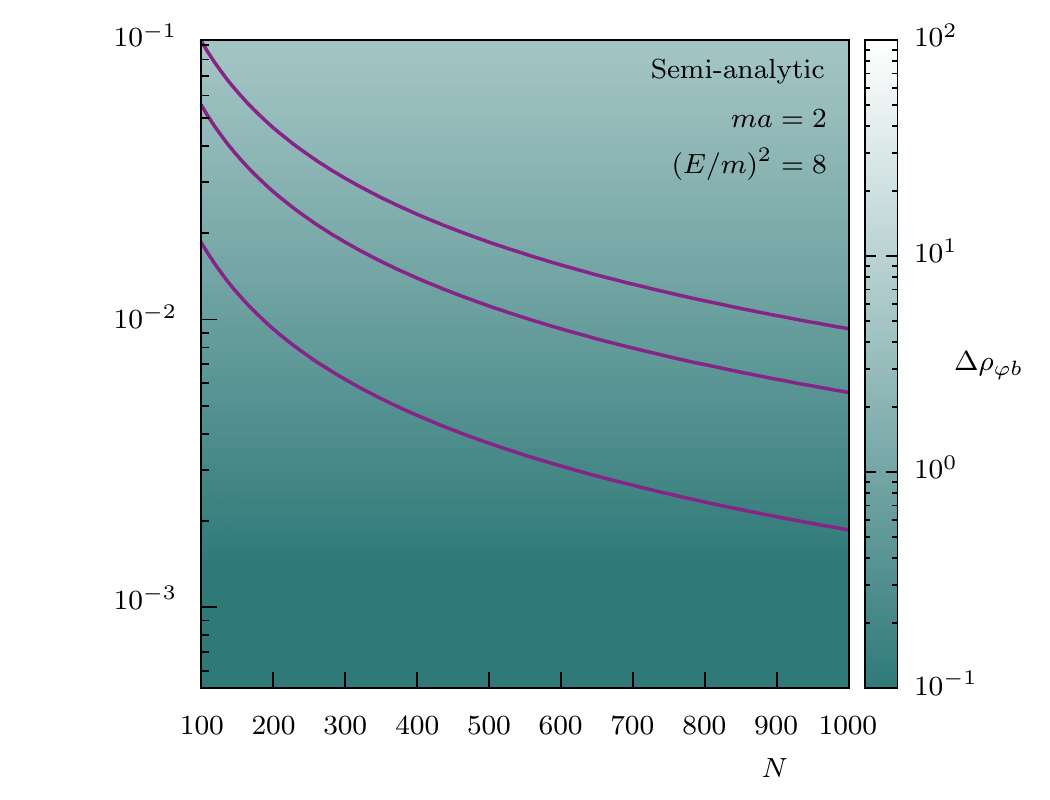}
 \caption{Density plot of $\Delta \rho_{\varphi b}$ as a function of $\epsilon / m^2$ and $N$ for a solution computed via the BF method (left) and the SA method (right) at fixed $ma = 2$ and $(E/m)^2 = 8$. Shown in red are the trajectories of constant $\eta = 2\pi N \epsilon_q / k_{\mathrm{max}}$ for $\eta = 5,$ 15, and 25.}
\label{fig:DelRho_map}
\end{figure}

In the $N\to\infty$ limit, any choice of $\eta$ satisfying the bound Eq.~\eqref{eq:eps_bound} should converge to the same solution.
In practice, we always work with finite $N$, and thus choices for $\eta$ lead to systematic effects. To minimize these systematic deviations, we explore an example solution as a function of both $\epsilon$ and $N$ for a given $ma$ and $E/m$. In Fig.~\ref{fig:DelRho_map}, we show a density plot of $\Delta \rho_{\varphi b}$ as a function of $\epsilon$ and $N$ for both the BF and SA methods at fixed $(E/m)^2 = 8$ and $ma = 2$ for $100\le N \le 1000$, along with curves at constant $\eta = 5,$ 15, and 25. For the BF method, for fixed $N$ and small enough $\epsilon$ the $\Delta \rho_{\varphi b}$ grows suddenly, meaning that we moved past the optimal $\epsilon(N)$ trajectory given by some optimal $\eta$, empirically verifying the approximate bound Eq.~\eqref{eq:eps_bound}. Moreover, we find clear signal of oscillating solutions for $\eta \lesssim 10$, which dampen as $N$ increases.

We find that for $\eta \sim 10$, there remains residual oscillations for the SA method as well, however the magnitude is considerably less than that of the BF and at the scale presented in Fig.~\ref{fig:DelRho_map} is indistinguishable.
Figure~\ref{fig:DelRho_map} also shows that for too large $\eta$, $\Delta \rho_{\varphi b}$ increases. 
Since we work with systems with $1000\le N \le 6000$, we choose to compute solutions with $10 \le \eta \le 25$ for both the BF and SA methods. We find that to our working precision, solutions computed with these $\eta$ yield consistent results. Deviations for different $\eta$ are an order of magnitude smaller than systematics estimated from the $N$ dependence of the solutions, as illustrated in Fig.~\ref{fig:a2_eta15_Nextra}, therefore we show results with $\eta = 15$ and absorb any $\eta$ fluctuations to the systematic errors arising from the $N$ dependence.

%%%%%%%%%%%%%
% FIGURE
%%%%%%%%%%%%%
%%%%%%%%%%%%%

%%%%%%%%%%%%%%%%%%%%%%%%%%%%%%%%%%%%
%	Section :: Above the three particle threshold
%%%%%%%%%%%%%%%%%%%%%%%%%%%%%%%%%%%%
\section{Above the three-particle threshold}
\label{sec:above_thr}

    %%%%%%%%%%%%%
    % FIGURE
    %%%%%%%%%%%%%
    \begin{figure}[t!]
        \centering
        \includegraphics[ width=0.95\textwidth]{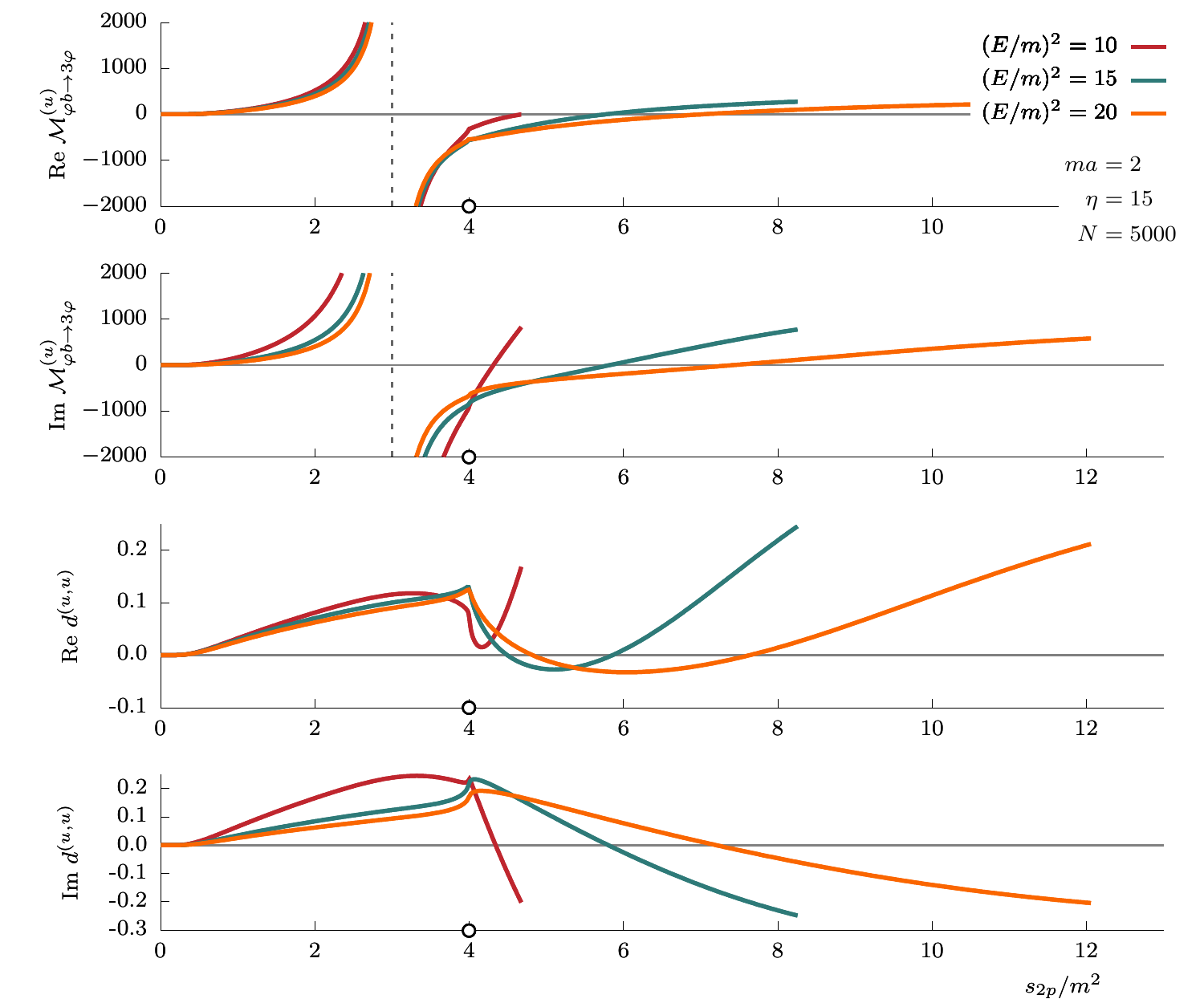}
        \caption{Real and imaginary parts of $\cM_{\varphi b\to 3\varphi}^{(u,u)}$ and $d^{(u,u)}$ amplitudes for fixed $ma = 2$, $\eta = 15$, and $N=5000$ as a function of $s_{2p}/m^2$ for three particle energies $(E/m)^2 = 10$, 15, and 20. The vertical dashed line in the top two panels shows the location of the two body bound state $s_b$ for this $ma$, and in all panels the two particle threshold at $s_{2p} = 4m^2$ is indicated by the open circle on the real axis.}
        \label{fig:Mphib3}
    \end{figure}
    %%%%%%%%%%%%%

So far, we have considered energies below the three-particle breakup threshold. In this section, we show that the integral equations can be solved above the three-particle threshold, which was a limitation of the method presented in Ref.~\citep{Romero-Lopez:2019qrt}. In this energy region, the bound state breakup amplitude $\varphi b \to 3\varphi$ is kinematically accessible. Following the discussion in Sec.~\ref{sec:IE_BS}, one obtains the $\varphi b\to 3\varphi $ amplitude via
    %%%%%
    \begin{align}
    \cM^{(u)}_{\varphi b,3\varphi}( p,E)
    =
    -\lim_{s_{2k}\to s_b}
    \frac{(s_{2k}- s_b)}{g}
    \, {\cM}^{(u,u)}_{3,S}(p, k).
    \end{align}
    %%%%%
Note, the final state is composed of three particles, and as a result the unsymmetrized amplitude emerges. 
In order to symmetrize it, one can follow a procedure similar to that of Eq.~\eqref{eq:symm_proc},
    %%%%%
    \begin{equation}
    \cM_{\varphi b,3\varphi}(E; \p, \mathbf{a}'  ) =  \sum_{\p \in \mathcal{P}_p }      \cM^{(u)}_{\varphi b,3\varphi}( p,E)\,,
    \end{equation}
    %%%%%
where only three terms are summed over, since one only needs to symmetrize the final state. Using similar arguments as in Sec.~\ref{sec:IE_BS}, we can write the $\varphi b \to 3 \varphi$ amplitude in terms of $d_S^{(u,u)}$,
    %%%%%
    \begin{align}
    \lim_{ \cK_\df\to 0}   \cM^{(u)}_{\varphi b,3\varphi}( p,E)
    &=
    g \, \cM_{2}(p) 
    \lim_{s_{2k}\to s_b}
    \, d^{(u,u)}_{S}( p,  k).
    \end{align}
    %%%%%
As before, we solve for $d_S^{(u,u)}$ for fixed initial $k = q$ at some $E$, only now the final state momentum $p$ is now free.

As an illustration of solving the integral equations above the three particle threshold, we show in Fig.~\ref{fig:Mphib3} the resulting $\cM^{(u)}_{\varphi b \to 3\varphi}$ and $d^{(u,u)}$ amplitudes as a function of $s_{2p}/m^2$ for fixed $k = q$, $ma = 2$, $\eta = 15$, $N = 5000$ at total energies $(E/m)^2 = 10$, 15, and 20. The two-body threshold behavior of $\cM^{(u)}_{\varphi b \to 3\varphi}$ at $s_{2p}/m^2 = 4$ is clearly visible and one can also see the final state bound state pole indicated by the dashed vertical line. Contrary to $\cM_{\varphi b \to 3\varphi }^{(u)}$, the $d^{(u,u)}$ amplitude does not exhibit this bound state pole, as originally discussed in Sec.~\ref{sec:IE}. The amplitudes are plotted within the allowed integration region $0 \le p \le k_{\mathrm{max}}(E)$, which in terms of $s_{2p}$ corresponds to $0 \le s_{2p}/m^2 \le (E/m-1)^2$.

%%%%%%%%%%%%%%%%%%%%%%%%%%%%%%%%%%%%
%	Section :: Conclusions
%%%%%%%%%%%%%%%%%%%%%%%%%%%%%%%%%%%%
\section{Conclusions}
\label{sec:conclusions}

In this work, we presented a numerical method for solving the three-body on-shell integral equations in the presence of the two-body bound-states. Our methodology approximates the integral equations as a system of $N$ linear equations, which are solved by usual matrix inversion techniques. The method is systematically improvable insofar as the mesh of momentum points that are used to generate the equations can be finer sampled, leading to larger systems that better converge to the $N\to \infty$ solution. Quantitatively, the quality of solutions is measured by computing the deviation from $S$ matrix unitarity for each $N$. Empirically we find the deviation $\Delta \rho_{\varphi b}$ decreases as $N$ increases, albeit slowly for matrix sizes of the order of $1000-6000$. In addition to finite $N$ effects, a convergence of solutions is affected by the presence of the two-body bound state, which produces a pole singularity in the region of integration.

We introduce two methods to circumvent the bound state pole. In the BF method, we regulate the pole singularity by introducing a finite $\epsilon$ which shifts the pole slightly off the real energy axis, and recover our solutions in the ordered limit of $\epsilon \to 0$ then $N\to \infty$. Alternatively, in the SA method, we remove the imaginary part of the pole explicitly using an $\epsilon$-regulated delta function. Using the NLO solution, we argued that the dominant systematic error is exponentially suppressed with $\eta \propto \epsilon N$ and we empirically verified this behavior. To ensure the ordered double limit is properly taken, we computed solutions for fixed $\eta$ and $N$. To increase the rate of convergence, and reduce the systematic error associated with finite $N$, we employed a strategy of extrapolating the results to the $N\to\infty$ limit by generating a sequence of solutions with various sizes $N$ and fitting the data to a model which is polynomial in $1/N$. The extrapolated solutions drastically reduce the deviation from unitarity by several orders of magnitude as compared to the finite $N$ results. As an assessment of the systematic error associated with this extrapolation, we generated a small ensemble of results by varying the included data in the fit. This is the dominating error of our studying, giving a sub-percent deviation from unitarity for the matrix sizes considered.

Our methodology is not limited to systems where the two-particle subsystem produces a bound state and can be adopted straightforwardly to resonating systems as well as processes with non-zero angular momenta. Additionally, the strategies presented here can be used to include the short-distance three-body $K$ matrix, which would be determined by complementary lattice QCD calculations. In principle, one may extend these methods to complex energies to systematically search below the threshold for three-particle bound states, or on unphysical sheets for resonances. However, it is currently unknown how to consistently analytically continue the solutions outside the physical region.

%%%%%%%%%%%%%%%%%%%%%%%%%%%%%%%%%%%%
%         Acknowledgements    
%%%%%%%%%%%%%%%%%%%%%%%%%%%%%%%%%%%%

\section{Acknowledgements}

RAB, AWJ, and SMD acknowledges support from U.S. Department of Energy contract DE-AC05-06OR23177, under which Jefferson Science Associates, LLC, manages and operates Jefferson Lab. RAB and MHI acknowledge support of the USDOE Early Career award, contract DE-SC0019229. SMD also acknowledges support by the U.S. Department of Energy Grant No. DE-FG02-87ER40365. The authors would like to thank J. Dudek, R. Edwards, M. Hansen, L. Leskovec, F. Romero-L\'opez, S. Sharpe, A. Szczepaniak, D. Wilson, and the rest of the Hadron Spectrum Collaboration for useful discussions.

%%%%%%%%%%%%%%%%%%%%%%%%%%%%%%%%%%%%
%	Bibliography
%%%%%%%%%%%%%%%%%%%%%%%%%%%%%%%%%%%%
\bibliography{main} %%% main.bib file
\end{document}